\documentclass[article]{aa}  
\usepackage{natbib}
\usepackage{graphicx}
\usepackage{txfonts}
\newcommand{\nH}  {\mbox{H}}                      % H 
\newcommand{\nD}  {\mbox{D}}                      % D 
 \newcommand{\HH}  {\mbox{H$_2$}}       %  H2
 \newcommand{\HD}  {\mbox{HD}}       %  HD
 \newcommand{\DD}  {\mbox{D$_2$}}       %  D2
\newcommand{\dC}  {\mbox{$^{12}$C}}       % 12C 
\newcommand{\tC}  {\mbox{$^{13}$C}}       % 13C  
\newcommand{\nCC}  {\mbox{C$_2$}}       % C2
\newcommand{\tCC}  {\mbox{$^{13}$CC}}       % 13CO
 \newcommand{\Cp}  {\mbox{$^{12}$C$^+$}}       % C+
\newcommand{\tCp}  {\mbox{$^{13}$C$^+$}}       % 13C+ 
\newcommand{\nO}  {\mbox{O}}       % O 
\newcommand{\dO}  {\mbox{$^{18}$O}}       % 18O 
\newcommand{\Op}  {\mbox{O$^+$}}       % O+ 
\newcommand{\nN}  {\mbox{$^{14}$N}}       %  N 
\newcommand{\fN}  {\mbox{$^{15}$N}}       % 15N 
\newcommand{\Np}  {\mbox{$^{14}$N$^+$}}       % 14N+
\newcommand{\fNp}  {\mbox{$^{15}$N$^+$}}       % 15N+
\newcommand{\nCO}  {\mbox{CO}}       %  CO
\newcommand{\tCO}  {\mbox{$^{13}$CO}}       % 13CO
\newcommand{\nCH}  {\mbox{CH}}       %  CH
\newcommand{\tCH}  {\mbox{$^{13}$CH}}       % 13CH
\newcommand{\nCD}  {\mbox{CD}}       %  CD
\newcommand{\nCN}  {\mbox{CN}}       %  CN
\newcommand{\nNH}  {\mbox{NH}}       % NH
\newcommand{\nND}  {\mbox{ND}}       % ND
\newcommand{\fNH}  {\mbox{$^{15}$NH}}       % 15NH
\newcommand{\fND}  {\mbox{$^{15}$ND}}       % 15ND
\newcommand{\tCN}  {\mbox{$^{13}$CN}}       % 13CN
\newcommand{\CfiN}  {\mbox{C$^{15}$N}}       %  C15N
 \newcommand{\NN}  {\mbox{N$_2$}}       %  N2
\newcommand{\NfiN}  {\mbox{N$^{15}$N}}       % N15N
\newcommand{\fiNN}  {\mbox{$^{15}$N$_2$}}       %  15N2
\newcommand{\nHCN}  {\mbox{HCN}}       % HCN
 \newcommand{\tHCN}  {\mbox{H$^{13}$CN}}       % H13CN
 \newcommand{\fHCN}  {\mbox{HC$^{15}$N}}       % HC15N
\newcommand{\tfHCN}  {\mbox{H$^{13}$C$^{15}$N}}       % H13C15N
\newcommand{\nHNC}  {\mbox{HNC}}       % HNC
\newcommand{\tHNC}  {\mbox{HN$^{13}$C}}       % HN13C 
\newcommand{\fHNC}  {\mbox{H$^{15}$NC}}       % H15NC
\newcommand{\nCCp}  {\mbox{C$_2$$^+$}}       % C2+
\newcommand{\fNNp}  {\mbox{$^{15}$NN$^{+}$}}       %  15NN+
 \newcommand{\NHp}  {\mbox{NH$^{+}$}}       %  NH+
\newcommand{\fNHp}  {\mbox{$^{15}$NH$^{+}$}}    %15NH+
\newcommand{\nNOp}  {\mbox{NO$^{+}$}}    %NO+
  \newcommand{\NDp}  {\mbox{ND$^{+}$}}       %  ND+
\newcommand{\fNDp}  {\mbox{$^{15}$ND$^{+}$}}       %  ND+
\newcommand{\NHHHp}  {\mbox{NH$_3 ^{+}$}}       %  NH3+
\newcommand{\NHDDp}  {\mbox{NHD$_2^{+}$}}       %  NHD2+
\newcommand{\NHHDp}  {\mbox{NH$_2$D$ ^{+}$}}       %  NH2D+
\newcommand{\NDDDp}  {\mbox{ND$_3 ^{+}$}}       %  ND3+
\newcommand{\NHHHHp}  {\mbox{NH$_4 ^{+}$}}       %  NH4+
\newcommand{\NHHHDp}  {\mbox{NH$_3$D$^{+}$}}       %  NH3D+
\newcommand{\NHHDDp}  {\mbox{NH$_2$D$_2^{+}$}}       %  NH2D2+
\newcommand{\NHDDDp}  {\mbox{NH D$_3^{+}$}}       %  NHD3+
\newcommand{\NDDDDp}  {\mbox{ND$_4 ^{+}$}}       %  ND4+
\newcommand{\NNHp}  {\mbox{N$_2$H$^{+}$}}       %  N2H+
 \newcommand{\NNDp}  {\mbox{N$_2$D$^{+}$}}       %  N2D+
 \newcommand{\fNNDp}  {\mbox{$^{15}$NND$^{+}$}}       %  N2D+
 \newcommand{\NfNDp}  {\mbox{N$^{15}$ND$^{+}$}}       %  N2D+
 \newcommand{\fNNHp}  {\mbox{$^{15}$NNH$^{+}$}}       %  15NNH+ 
 \newcommand{\NfNHp}  {\mbox{N$^{15}$NH$^{+}$}}       %  N15NH+
 \newcommand{\ffNNHp}  {\mbox{$^{15}$N$_2$H$^{+}$}}       %  N15NH+
 \newcommand{\nNO}  {\mbox{$^{14}$NO}}       %  NO
 \newcommand{\fNO}  {\mbox{$^{15}$NO}}       %  15NO
 \newcommand{\HCNHp}  {\mbox{HCNH$^{+}$}}       %  HCNH+
 \newcommand{\HCNDp}  {\mbox{HCND$^{+}$}}       %  HCND+
 \newcommand{\DCNHp}  {\mbox{DCNH$^{+}$}}       %  DCNH+
 \newcommand{\HtCNHp}  {\mbox{H$^{13}$CNH$^{+}$}}       %  H13CNH+
 \newcommand{\HCfNHp}  {\mbox{HC$^{15}$NH$^{+}$}}       %  HC15NH+
\newcommand{\NHHH}  {\mbox{NH$_3$}}       % NH3
\newcommand{\NHH}  {\mbox{NH$_2$}}       % NH2
\newcommand{\NHHD}  {\mbox{NH$_2$D}}       % NH2D
\newcommand{\fNHHH}  {\mbox{$^{15}$NH$_3$}}       % 15NH3
 \newcommand{\fNHHD}  {\mbox{$^{15}$NH$_2$D}}       % 15NH2D
 \newcommand{\CNCp}  {\mbox{CNC$^{+}$}}       %  CNC+
 \newcommand{\CCN}  {\mbox{C$_2$N}}       %  C2N
 \newcommand{\CCfN}  {\mbox{C$_2$$^{15}$N}}       %   C215N 
 \newcommand{\fNCCN}  {\mbox{$^{15}$NCCN}}       %   15NCCN 
 \newcommand{\CfNCp}  {\mbox{C$^{15}$NC$^{+}$}}       %   C15NC+
 \newcommand{\NOp}  {\mbox{NO$^{+}$}}       %  NO+
 \newcommand{\COp}  {\mbox{CO$^{+}$}}       %  CO+
  \newcommand{\HCOp}  {\mbox{HCO$^{+}$}}       %  HCO+
 \newcommand{\HtCOp}  {\mbox{H$^{13}$CO$^{+}$}}       %  H13CO+
 \newcommand{\DCOp}  {\mbox{DCO$^{+}$}}       %  DCO+
 \newcommand{\HHHp}  {\mbox{H$_{3}$$^{+}$}}       %  H3+
 \newcommand{\HHDp}  {\mbox{H$_{2}$D$^{+}$}}       %  H2D+
\newcommand{\cm}{cm$^{-1}$}

\begin{document}

   \title{Isotopic fractionation of carbon, deuterium and nitrogen: a full chemical study}

 %\subtitle{I. revisiting the chemical mechanisms}

   \author{E. Roueff
          \inst{1,2}%\inst{2}
          \and
          J.C. Loison\inst{3}%\fnmsep \thanks{Just to show the usage
          %of the elements in the author field}
          \and
          K.M. Hickson\inst{3}
          }

   \institute{LERMA, Observatoire de Paris, PSL Research University, CNRS, UMR8112, Place Janssen, 92190 Meudon Cedex, France\\
              \email{evelyne.roueff@obspm.fr}
         \and
            Sorbonne Universit\'es, UPMC Univ. Paris 6
          \and
            ISM, Universit\'e de Bordeaux - CNRS, UMR 5255, 351 cours de la Lib\'eration, 33405 Talence Cedex, France\\
             \email{jean-christophe.loison@u-bordeaux.fr}
          %   \thanks{The university of heaven temporarily does not
              %       accept e-mails}
             }

   \date{Received     ; accepted     }

% \abstract{}{}{}{}{} 
% 5 {} token are mandatory
 \abstract
   {The increased sensitivity and high spectral resolution of millimeter telescopes allow  the detection of an increasing number of isotopically substituted molecules in the interstellar medium.  The \nN / {\fN}  ratio is 
 difficult to measure directly for carbon containing molecules.  }
   {  We want to check the underlying hypothesis that the \tC / {\dC} ratio of nitriles and isonitriles   is equal to the elemental value via a chemical time dependent gas phase chemical model.}
   {We have built a chemical network containing D, {\tC} and {\fN} molecular species after a careful check of the possible fractionation reactions at work in the gas phase. }
   {Model results obtained for 2 different physical conditions corresponding respectively to a moderately dense cloud in an early evolutionary stage and a dense depleted pre-stellar core tend to show that ammonia and its singly deuterated form are somewhat enriched in \fN, in agreement with observations. The \nN / {\fN} ratio in {\NNHp} is found to be close to the elemental value, in contrast to previous models which obtain a significant enrichment, as we found that the fractionation reaction between {\fN} and {\NNHp} has a barrier in the entrance channel. The large values  of  the { \NNHp /\fNNHp} and  { \NNHp / \NfNHp}   ratios derived  in L1544 cannot be reproduced in our model. Finally we find that nitriles  and isonitriles  are in fact significantly  depleted in {\tC}, questioning previous interpretations of observed C\fN, {\fHCN} and {\fHNC} abundances from {\tC} containing isotopologues.}
   {}


   \maketitle
%
%________________________________________________________________

\section{Introduction}
\label{sec:intro}
Understanding isotopic abundances on a large scale is a major  field of interest which has received a great deal of attention for its application to terrestrial environments (ocean, meteorites), the solar system (planets, comets) and in galactic interstellar space. The variation of isotopic ratios may give us some information about the link between solar system objects and galactic interstellar environments as discussed by \cite{aleon:10}.
 We principally focus our study on interstellar environments where low temperature conditions may significantly impact the isotopic ratios of the molecular content. Isotopic molecules are detected in a variety of environments and offer an additional tool to determine physical conditions as they usually do not suffer from opacity problems.
 
 Early modeling studies on {\tC} and {\dO} isotopic enrichment were performed by \cite{langer:84} who introduced  different isotopic exchange reactions, relying on previous theoretical and experimental studies by \citealt{watson:76,smith:80}. Possible effects of selective photodissociation of CO have  subsequently been emphasized \citep{glassgold:85,lebourlot:93,visser:09}, which tend to increase the {\nCO/\tCO}  ratio. The use of the CN radical as a tracer of  {\dC/\tC} isotopic ratio has been raised by \cite{savage:02,milam:05}, who studied the corresponding gradient as a function of the galactic distance. The actual value of the {\dC/\tC} isotopic ratio in the local InterStellar Medium (ISM) is assumed to be 68 \citep{milam:05}.
  
 The possibility of nitrogen isotopic fractionation in interstellar clouds has been investigated by \cite{terzieva:00} who suggested various {\fN} isotopic exchange reactions. \cite{rodgers:04,rodgers:08b,rodgers:08a} used these suggested reaction rate constants to  predict nitrogen isotopic fractionation in chemical models of dense interstellar molecular cores. They specifically discussed the role of the atomic to molecular nitrogen ratio in the fractionation process and  the possible link between nitrogen hydrides and CN containing molecules (nitriles).
The corresponding observations are sparse however and difficult as the elemental {\nN/\fN} ratio is high (441 $\pm$ 5), as determined by the recent Genesis solar wind sampling measurement \citep{marty:11} and assumed to hold in the local ISM. In addition, the zero point energy (ZPE) differences involved in nitrogen fractionation reactions are small and the predicted corresponding chemical enrichment is moderate. 
\cite{lucas:98} reported {\fHCN} absorption in diffuse clouds located in front of distant quasars with a {\nHCN/\fHCN} ratio of 270$\pm$ 27, close to value reported for the Earth. However, various new observations of isotopic nitrogen containing molecules have been reported, including {\fN}  substituted ammonia and deuterated ammonia \citep{gerin:09,lis:10} , the  diazenylium  ion (\NNHp)  \citep{bizzocchi:10,bizzocchi:13,daniel:13}, CN and HCN \citep{ikeda:02,pillai:07,adande:12,hilyblant:13,daniel:13}, HCN and HNC \citep{wampfler:14}.  The strong depletion found in  {\fN}  variants of the {\NNHp} isotopologue strongly contradicts model predictions \citep{gerin:09}, which motivates a reinvestigation of the chemical processes at work.  With this in mind, the link between deuterated chemistry and the possible role of ortho/para molecular hydrogen has been 
studied by \cite{wirstrom:12}. 
 
We analyse in Section~\ref{sec:reac} the various possible isotopic exchange reactions that are involved for carbon and nitrogen containing molecules. Indeed, most nitrogen fractionation observational results for CN containing molecules involve only {\tC} and
{\fN} species so that the measure of the nitrogen isotopic ratio  assumes a  fixed {\dC/\tC} fraction. We examine and extend the pioneering study of \cite{terzieva:00} %in  updating the zero point energy (ZPE) values involved in the isotopic exchange reactions, and in the absence of any experimental information,% perform  DFT calculations based on the analysis of intermolecular interactions. We specifically
%discuss the liability of a barrier in the entrance channel and take advantage, as much as possible, of experimental information.
and check for the possible presence of barriers in the entrance channels of isotopic exchange reactions  through theoretical calculations. 
% through theoretical DFT (Density Functional Theory) calculations.
We also update the zero point energy (ZPE) values involved   and derive the corresponding exothermicity values.
We present our new chemical model in Section~\ref{sec:model} and compare with available observations and other models. Our conclusions are presented in Section~\ref{sec:conclusion}

\section{Chemical reactions involving isotopic substitutes of {\dC} and \nN.}
\label{sec:reac}
\subsection{{\tC} and {\fN} exchange reactions}
At very low temperatures, isotopic exchange reactions may only occur  if no barrier is present between the interacting atoms, ions and molecules or if tunnelling plays an important role. Experimental information is crucial and we constrain the evaluation of rate constants using that information. If no experimental data are available, we apply theoretical  methods to determine the presence of a barrier: A first technique uses
 DFT (Density Functional Theory)  calculations (with the hybrid M06-2X functional developed by \cite{zhao:08} which is well suited for thermochemical calculations, associated to the cc-pVTZ basis set using GAUSSIAN09 software). The alternative is provided by the MRCI+Q method (with the aug-cc-pVTZ basis set).
For barrierless cases, we derive the reaction rate constants  by using a simple capture theory \citep{georgievskii:05}  for both ion-neutral and neutral-neutral reactions. %We also follow the general rule that isotopic exchange does not take place
%when another exothermic reactive channel is present.
We consider four different families of isotopic exchange reactions:
\begin{itemize}
\item{{\emph{A : direct reactions}}. The proton transfer in the $\NNHp + \NfiN \rightarrow  \NfNHp + \NN$ reaction can serve as an example.  In this case and for reactions without a barrier, the reaction rate coefficient of the forward reaction is equal to the capture rate constant multiplied by a probability factor {\it{f(B, M)}}, $f(B,M)$ depending on the rotational constant, mass and symmetry values of the reactants and products. In reactions involving{\fN} and {\tC} the mass ratio of reactants and products are very close and $f(B,M) \cong \sigma_{\rm{entrance~channel}} / \sigma _{\rm{exit~channels} }$} (The symmetry number $\sigma$ is equal to the number of pathways). The reverse reaction is calculated from the equilibrium constant $K$, as in \cite{terzieva:00}: $K = k_f/k_r=f(B,M) \times exp( \Delta E / kT)$.
 
\item{{\emph{B : reactions involving adduct formation leading to direct products without isomerization}}. As an example, we refer to 
$\fNp + \NN \rightarrow  \NfiN  + \Np$.  We first assume that the high pressure rate constant is equal to the capture rate constant (for reactions without a barrier).  We apply statistical theory  for the system at thermal equilibrium so that $ k_f + k_r = k_{capture} $. From the equilibrium constant expression, we then derive 
$ k_f=  k_{capture} \times \frac{f(B,M)}{[ f(B,M) + exp(- \Delta E / kT)] }$ and  $ k_r=  k_{capture} \times \frac{exp(- \Delta E / kT)}{[f(B,M) +exp(- \Delta E / kT)]}$} 

\item{{\emph{C : reactions involving adduct formation with  isomerization pathways}}. Such a case holds for 
$\tC + \nHCN  \rightarrow  \dC  + \tHCN$. We again assume that the high pressure rate constant is given by capture theory (for reactions without a barrier). The isotopic isomerization reaction competes with the dissociation of the adduct. The rate constant depends on the location of the transition state and statistical calculations are generally required to estimate the isomerization reaction rate constant.}
 \item{{\emph{D : other reactive exothermic channels exist}}}. The exchange reaction is   discarded generally (the possibility of  N atom exchange in the {\fN} + CN and in {\fN} + {\CCN} reactions is discussed).
\end{itemize}
The knowledge of the exoergicity values $\Delta E$ is  also a major concern. They are obtained from the differences of  the ZPEs between products and reactants. We recall in the Appendix  the corresponding expressions and derive their  values by using the most recent determinations of spectroscopic constants.% of different nitrogen and carbon containing molecules and corresponding ZPEs, which will be used in the discussion in Section~\ref{sec:rr}.  
 
 We summarize in Table \ref{tab:exch}  the different isotopic exchange reactions considered and display the corresponding reaction rate constants. Detailed information is provided in the online material on the theoretical methods used for the different  systems. The reactions involving {\fN} are displayed in the upper part of the Table. We also consider {\tC} isotopic exchange reactions in the lower part of Table \ref{tab:exch}.  
 \begin{table*}[h]
\caption{Isotopic exchange reactions.}             
\label{tab:exch}      
\begin{center}          
\begin{tabular}{l l l l l c  l}     % 7 columns 
\hline\hline       
                      % To combine 4 columns into a single one 
%HJD & $E$ & Method\#2 & \multicolumn{4}{c}{Method\#3}\\ 
label /  & \multicolumn{3}{c}{Reaction}  &   k$_f$  $^*$ & $f(B,M)$ $^*$  & $\Delta$E  $^*$  \\
comment  &  \multicolumn{3}{c}{ }  &     (cm$^{3}$ s$^{-1}$) &   &   (K)\\
\hline                    
(1)  A  & {\bf{$\NfiN + \NNHp$}} & $\rightleftharpoons$ &  {\bf{$ \NfNHp + \NN$}} &  2.3 $\times$ 10$^{-10}$ &  0.5  &  10.3   \\
 (2) A &{\bf{$\NfiN + \NNHp$}} & $\rightleftharpoons$ &  {\bf{$ \fNNHp + \NN$}}  &       2.3 $\times$ 10$^{-10}$   & 0.5   &   2.1   \\
 (3) A &{\bf{$\NfiN + \fNNHp$}} & $\rightleftharpoons$ &  {\bf{$ \NfNHp + \NfiN$}}  &       4.6 $\times$ 10$^{-10}$   & 1  &   8.1   \\
(4)  B   &  {\bf{$\fNp + \NN$}} & $\rightleftharpoons$ &  {\bf{$ \Np + \NfiN$}} & 4.8 $\times  10 ^{-10} \times   \frac{2}{2+exp(-28.3/T)}$ &2 & 28.3   \\
(5)  C  & {\bf{$\fN  + \CNCp$}} & $\rightleftharpoons$ &      {\bf{$ \CfNCp  + \nN$}} & 3.8 $\times  10 ^{-12} \times  (\frac{T}{300})^{-1} $  &   1  & 38.1  \\ 
(6)  D &  {\bf{$\fNp  + \nNO$}} & $\rightleftharpoons$ &      {\bf{$ \Np + \fNO$}} &   no react   &   -     &  24.3  \\
(7) barrier   & {\bf{$\fN  + \NNHp$}} & $\rightleftharpoons$ &      {\bf{$\nN  +   \NfNHp $}} &    no react &  -  &  38.5 \\ 
(8)  barrier &   {\bf{$\fN  + \NNHp$}} & $\rightleftharpoons$ &      {\bf{$\nN  +   \fNNHp $}} &  no react  &  -  &  30.4 \\ 
(9)  barrier &   {\bf{$ \fNNHp + \nH  $}} & $\rightleftharpoons$ &      {\bf{$\nH  +   \NfNHp $}} &  no react   &  -  &  8.1 \\ 
 (10)   barrier  & {\bf{$\fN  + \HCNHp$}} & $\rightleftharpoons$ &      {\bf{$\nN  +   \HCfNHp $}} &  no react  & -   & 37.1  \\ 
 (11)  D & {\bf{$\fN  + \nCN$}}  &  $\rightleftharpoons$      &   {\bf{$ \nN  + \CfiN$}} &  upper limit  : 2.0 $\times$ 10$^{-10}$    $\times$ &  1   & 22.9   \\ 
                                 &     &        &     &  (T/300)$^{1/6}$  $\times$ $\frac{1}{1+exp(-22.9/T)}$   &   &  \\
  (12)  B & {\bf{$\fN  + \CCN$}} &  $\rightleftharpoons$ &      {\bf{$ \nN  + \CCfN $}} &     1.6 $\times  10 ^{-10} \times  (T/300)^{1/6} \times$ &  1  &   26.7  \\ 
                                 &     &        &     &  $\frac{1}{1+exp{\bf{(-26.7/T)}}}$   &   &  \\
(13)  D &  {\bf{$\fN   + \nNO$}} & $\rightleftharpoons$ &      {\bf{$\nN  + \fNO$}} &     -    &   -     &  24.3  \\ 
\hline                    
\hline                    
(14)   B &     {\bf{$\tCp  + \nCO$}} & $\rightleftharpoons$ &      {\bf{$ \Cp  + \tCO$}} & 6.6 $\times  10 ^{-10} \times (T/300)^{-0.45} $&  1   &   34.7   \\
        &      &         &      &    $\times$  exp(-6.5/T)    $  \times$  $ \frac{1}{1+exp(-34.7/T)}$  &     &      \\
(15)  A  &  {\bf{$\tCO   + \HCOp$}} & $\rightleftharpoons$ &      {\bf{$ \nCO   + \HtCOp$}} & 2.6 $\times  10 ^{-10}  \times (T/300)^{-0.4} $  & 1 &  17.4 \\ 
(16)   B & {\bf{$\tCp  + \nCN$}} & $\rightleftharpoons$ &      {\bf{$ \Cp  + \tCN$}} &   3.82 $\times  10 ^{-9} \times (T/300)^{-0.4}$    &  1   &   31.1   \\ 
           &   &       &      &   $ \times$   $ \frac{1}{1+exp(-31.1/T)}$  &      &      \\ 
 (17)  B  &    {\bf{$\tC   + \nCN$}} & $\rightleftharpoons$ &      {\bf{$ \dC   + \tCN$}} &  3.0 $\times  10 ^{-10}   \times \frac{1}{1+exp(-31.1/T)}$  &  1   &   31.1   \\ 
 (18)  C & {\bf{$\tC   + \nHCN$}} & $\rightleftharpoons$ &      {\bf{$ \dC   + \tHCN$}} & no react  & - &  48.4 \\ 
(19)  B &  {\bf{$\tC   + \nCC$}} & $\rightleftharpoons$ &      {\bf{$ \dC   + \tCC$}} &   3.0 $\times  10 ^{-10}  \times  \frac{2}{2+exp(-26.4/T)}$  &  2   &   26.4   \\ 
  (19) barrier &{\bf{$ \tCH + \nCO $}} & $\rightleftharpoons$ &      {\bf{$ \tCO   + \nCH$}} & no react & - &   28.6 \\ 
  \hline                  
\end{tabular}
\end{center}
$^*$    k$_f$ is the forward reaction rate coefficient. The reverse reaction rate coefficient,   k$_r$ , is obtained by   k$_r$ = $\frac{k_f}{f(B,M)}$ exp(- $\Delta$E/T).
\end{table*}
Table \ref{tab:exch} shows two main discrepancies compared to previous studies by \cite{terzieva:00} : The  exchange reactions between atomic {\fN}  and {\NNHp},  {\HCNHp} are found to be unlikely to occur as significant barriers arise in the complex formation step. A similar result is obtained for {\fNp} exchange with NO, whereas these reactions had been included in \cite{terzieva:00}. The exchange reaction between atomic {\fN} and CN, which was suggested by \cite{rodgers:08b}, is found to be plausible.
Additional possibilities of exchange have also been considered such as the reaction between {\fN}  and {\CCN}. As far as {\tC} possible fractionation is concerned, we find that CN could be enriched in  {\tC} through the exchange reactions of CN with {\tC} and {\tCp}. However, such a mechanism does not hold for HNC as atomic carbon is found to react with HNC \citep{loison:14}. $^{13}$C
enrichment of HCN is also found to be unlikely as the calculated transition state lies above the entrance level in the hypothetical isomerization process (C mechanism). 

 \subsection{Ammonia synthesis}
 Ammonia synthesis proceeds mainly through a chain of reactions starting with {\Np} and {\HH}, as the reaction between N and {\HHHp} has been shown to be inefficient \citep{milligan:00}.
 \label{sec:ammonia}
 \subsubsection{The {\Np}  +  {\HH}   reaction and isotopic substitutions}
\label{sec:nph2}
This almost thermoneutral reaction deserves a special mention and has received  considerable attention. \cite{lebourlot:91} first pointed out the possible role of ortho-{\HH} in the interstellar chemistry of ammonia as the energy of its J=1 rotational level almost compensates the small endothermicity of the reaction $ \Np   +   \HH  \rightarrow \NHp + \nH$. \cite{dislaire:12}  subsequently  reanalysed the experimental data \citep{marquette:88} and suggested new separate expressions for the reaction rate with {p-\HH} and o-\HH. Similar results were obtained by \cite{zymak:13}, who also emphasized the possible role of the fine structure level of \Np.  We follow the prescription derived by \cite{dislaire:12} and  extend their analysis to deuterated forms and those including {\fN} , as displayed in Table \ref{tab:npHH}.   In the case of {\fN}  substituted compounds, we have taken into account the (small) additional term due to the change in  ZPE. 
\begin{table*}[h]
\caption{Reaction rate coefficients of   {\mbox{N$^{+}$}}   +  {\HH}  and isotopic variants.}             
\label{tab:npHH}      
\centering          
\begin{tabular}{l c l l c }     % 5 columns 
\hline\hline       
                      % To combine 4 columns into a single one 
%HJD & $E$ & Method\#2 & \multicolumn{4}{c}{Method\#3}\\ 
\multicolumn{3}{c}{Reaction}  &   k   (cm$^{3}$ s$^{-1}$)& Comment\\
\hline                    
{\bf{$\Np  + p-\HH$}} & $\rightarrow$ &      {\bf{$ \NHp  + \nH$}}& 8.35 $\times$ 10$^{-10}$ $\times$ exp(-168.5/T)  & \citealt{dislaire:12}\\  
{\bf{$\Np  + o-\HH$}} & $\rightarrow$ &      {\bf{$ \NHp  + \nH$}}& 4.2 $\times$ 10$^{-10}$ $\times$ (T/300)$^{-0.17}$ $\times$ exp(-44.5/T) & \citealt{dislaire:12} \\  
{\bf{$\fNp  + p- \HH$}} & $\rightarrow$ &      {\bf{$ \fNHp  + \nH$}}& 8.35 $\times$ 10$^{-10}$$\times$ exp(-164.3/T)&  see text  \\  
{\bf{$\fNp  + o- \HH$}} & $\rightarrow$ &      {\bf{$ \fNHp  + \nH$}}& 4.2 $\times$ 10$^{-10}$ $\times$ (T/300)$^{-0.17}$ $\times$ exp(-39.7/T)  & see text \\  
{\bf{$\Np  + \HD   $}} & $\rightarrow$ &      {\bf{$ \NDp  + \nH$}}& 3.17 $\times$ 10$^{-10}$  $\times$ exp(-16.3/T) & \citealt{marquette:88}  \\  
{\bf{$\Np  + \HD   $}} & $\rightarrow$ &      {\bf{$ \NHp  + \nD$}}& 3.17 $\times$ 10$^{-10}$  $\times$ exp(-594.3/T) & see text\\  
{\bf{$\fNp  + \HD   $}} & $\rightarrow$ &      {\bf{$ \fNDp  + \nH$}}& 3.17 $\times$ 10$^{-10}$  $\times$ exp(-9.3/T)  & see text \\  
{\bf{$\fNp  + \HD   $}} & $\rightarrow$ &      {\bf{$ \fNHp  + \nD$}}& 3.17 $\times$ 10$^{-10}$  $\times$ exp(-589.5/T) & see text \\  
{\bf{$\Np  + \DD   $}} & $\rightarrow$ &      {\bf{$ \NDp  + \nD$}}& 2.37 $\times$ 10$^{-10}$  $\times$ exp(-197.9/T)  &  \citealt{marquette:88} \\  
{\bf{$\fNp  + \DD   $}} & $\rightarrow$ &      {\bf{$ \fNDp  + \nD$}}& 2.37 $\times$ 10$^{-10}$  $\times$ exp(-190.9/T) &  \citealt{marquette:88}  \\  
 \hline                  
\end{tabular}
\end{table*}
 These expressions should be used with caution as in their kinetic expression, we consider that the exponential term represents the enthalpy difference between the products and reactants. The capture rate constant of  {\fNp} + HD reaction is about (2/3)$^{0.5}$
smaller than the rate of the {\fNp} + {\HH} reaction, due to the different mass dependences. The formation of {\fNDp} is favored at low temperatures. %The $f$  factor becomes 0.5 for 
%{\fNp} + {\DD} at high temperature.
%
\subsubsection{ {\NHHHp} + \HH}
The final step of ion-molecule reactions leading to ammonia formation is the reaction between {\NHHHp} and \HH,  giving \NHHHHp. Although exothermic, this reaction has a strong temperature dependence, displaying a minimum at T $\sim$ 100K and a slow increase at lower  temperatures \citep{barlow:87}, which is interpreted by the presence of a barrier to complex formation, as discussed in \cite{herbst:91}. At temperatures close to 10K, the reaction is likely to proceed through tunneling which may take place or H atom  abstraction. However, the {\NHHHp} reaction with D$_2$ is found to be slower when the temperature decreases as tunneling is not efficient  with deuterium. We thus reconsider the isotopic variants of this reaction, as shown in Table \ref{tab:nhhhp}, where we give the present reaction rates compared to previous values which were derived from \cite{anicich:86}.
These values are indeed different from  those used in our previous studies \citep{roueff:05} where we assumed the same rate for  the channels resulting from the reactions of {\NHHHp} and isotopologues with HD, based on pure statistical considerations. 
 We discuss the resulting modifications in Section \ref{sec:model}.
\begin{table*}[h]
\caption{Reaction rate coefficients of   {\NHHHp}   +  {\HH}  and isotopic variants at T = 10K.}             
\label{tab:nhhhp}      
\begin{center}         
\begin{tabular}{l c l c c}     % 5 columns 
\hline\hline       
                      % To combine 4 columns into a single one 
%HJD & $E$ & Method\#2 & \multicolumn{4}{c}{Method\#3}\\ 
\multicolumn{3}{c}{Reaction}  &   \multicolumn{2}{c}{k   (cm$^{3}$ s$^{-1}$)} \\
    &    &   &   present work & old value (*)  \\
\hline                    
{\bf{$\NHHHp  +  \HH$}} & $\rightarrow$ &      {\bf{$ \NHHHHp  + \nH$}}& 8.2 $\times$ 10$^{-13}$  &  2.4 $\times$ 10$^{-12}$ \\  
 {\bf{$\NHHHp  +  \HD$}} & $\rightarrow$ &      {\bf{$ \NHHHHp  + \nD$}}& 8.2 $\times$ 10$^{-13}$ &  1.2 $\times$ 10$^{-12}$  \\  
 {\bf{$\NHHHp  +  \HD$}} & $\rightarrow$ &      {\bf{$ \NHHHDp  + \nH$}}& 1.0 $\times$ 10$^{-13}$  &  1.2 $\times$ 10$^{-12}$ \\  
{\bf{$\NHHHp  +  \DD$}} & $\rightarrow$ &      {\bf{$ \NHHHDp  + \nD$}}& 1.0 $\times$ 10$^{-13}$  &  2.4 $\times$ 10$^{-12}$ \\  
 {\bf{$ \NHHDp  + \HH$}} & $\rightarrow$ &      {\bf{$ \NHHHDp  + \nH$}}& 8.2 $\times$ 10$^{-13}$  &  2.4 $\times$ 10$^{-12}$ \\  
 {\bf{$ \NHHDp  + \HD$}} & $\rightarrow$ &      {\bf{$ \NHHHDp  + \nD$}}& 8.2 $\times$ 10$^{-13}$  & 1.2 $\times$ 10$^{-12}$  \\  
 {\bf{$ \NHHDp  + \HD$}} & $\rightarrow$ &      {\bf{$ \NHHDDp  + \nH$}}& 1.0 $\times$ 10$^{-13}$  & 1.2 $\times$ 10$^{-12}$  \\  
{\bf{$ \NHHDp  + \DD$}} & $\rightarrow$ &      {\bf{$ \NHDDDp  + \nH$}}& 1.0 $\times$ 10$^{-13}$  &  2.4 $\times$ 10$^{-12}$ \\  
 {\bf{$ \NHDDp  + \HH$}} & $\rightarrow$ &      {\bf{$ \NHHDDp  + \nH$}}& 8.2 $\times$ 10$^{-13}$   &  2.4 $\times$ 10$^{-12}$ \\  
 {\bf{$ \NHDDp  + \HD$}} & $\rightarrow$ &      {\bf{$ \NHHDDp  + \nD$}}& 8.2 $\times$ 10$^{-13}$  &  1.2 $\times$ 10$^{-12}$ \\  
 {\bf{$ \NHDDp  + \HD$}} & $\rightarrow$ &      {\bf{$ \NHDDDp  + \nH$}}& 1.0 $\times$ 10$^{-13}$  & 1.2 $\times$ 10$^{-12}$  \\  
{\bf{$ \NHDDp  + \DD$}} & $\rightarrow$ &      {\bf{$ \NHDDDp  + \nD$}}& 1.0 $\times$ 10$^{-13}$  & 2.4 $\times$ 10$^{-12}$  \\  
{\bf{$\NDDDp  +  \HH$}} & $\rightarrow$ &      {\bf{$ \NHDDDp  + \nH$}}& 8.2 $\times$ 10$^{-13}$  & 2.4 $\times$ 10$^{-12}$  \\  
 {\bf{$\NDDDp  +  \HD$}} & $\rightarrow$ &      {\bf{$ \NHDDDp  + \nD$}}& 8.2 $\times$ 10$^{-13}$ &   1.2 $\times$ 10$^{-12}$ \\  
 {\bf{$\NDDDp  +  \HD$}} & $\rightarrow$ &      {\bf{$ \NDDDDp  + \nH$}}& 1.0 $\times$ 10$^{-13}$   &  1.2  $\times$ 10$^{-12}$ \\  
{\bf{$\NDDDp  +  \DD$}} & $\rightarrow$ &      {\bf{$ \NDDDDp  + \nD$}}& 1.0 $\times$ 10$^{-13}$  &   2.4 $\times$ 10$^{-12}$  \\  
  \hline                  
\end{tabular}
\end{center}
(*) \cite{roueff:05}
\end{table*}
Identical reaction rate coefficients  are used for the $^{15}$N isotopically substituted reactions. 
 \section{Models}
\label{sec:model}
\subsection{General features}
Chemical reactions involving nitrogen atoms and CH, CN and OH have  been   studied experimentally at low temperatures and shown to be less efficient than previously thought at low temperatures \citep{daranlot:12,daranlot:13}, which was confirmed by theoretical studies. %This induces a higher atomic / molecular nitrogen ratio in interstellar cloud models. 
The corresponding reaction rate constants have been implemented in the KIDA chemical data base \citep{wakelam:13} and we have updated our chemical network accordingly. We also include the reactions discussed in \cite{loison:14} in their study of HCN / HNC chemistry. We  explicitly include deuterium, {\tC} and {\fN} molecular compounds in our chemical network. The reactions displayed  in Table \ref{tab:exch}, have been included,  which allows us to test the hypothesis of a constant {\dC}/{\tC} isotopic ratio to derive the {\nN/\fN} ratio in {\CfiN} containing molecules. 
We take into account the role of the ortho/para ratio of  molecular {\HH}  in the  {\HHDp} + {\HH} and {\Np} ({\fNp}) + {\HH} chemical reactions   in an approximate way: we do not compute the full ortho/para equilibrium  as in  the models of \citealt{flower:06,pagani:11,faure:13} and rather introduce it as a model parameter, which can be varied.  \cite{faure:13} have found a value of 10$^{-3}$    for temperatures below 15K.%, which has been used in the models described below. 

Apart from exchange reactions, reactions involving isotopic molecules are assumed to have the same total rate constant as those involving the main isotope except for the reaction of {\fNp}  with H$_{2}$/HD/D$_{2}$. The various reaction channels  are obtained from statistical considerations, in the absence of experimental information. We restrict carbon containing molecules to 3 carbon atoms, nitrogen containing molecules to 2 nitrogen atoms and consider full deuteration as in our previous studies \citep{roueff:05}. Within these constraints, the number of species considered in the model is 307 linked through more than 5400 chemical reactions. 
\begin{table}[h]
\caption{Model definitions. The { C / \tC}      and  {N / \fN}  ratios are respectively taken as 68 \citep{milam:05} and 440
 \citep{marty:11}.  }   
\label{tab:model}      
\centering          
\begin{tabular}{lll }     % 3columns 
\hline\hline       
     & Model (a) &    Model (b)  \\
     \hline
density {\mbox{n$_H$}} (cm$^{-3}$) &    2 $\times$ 10$^4$ &   2 $\times$ 10$^5$  \\
Temperature (K) &  10  &  10 \\
cosmic ionization rate per H$_2$ (s$^{-1}$) &  1.3  $\times$ 10$^{-17}$     &   1.3 $\times$ 10$^{-17}$      \\
%o/p H$_2$ &   10$^{-3}$  &  10$^{-3}$  \\
\hline
He {/} H  & 0.1  &  0.1     \\ 
C {/} H &   4.15 $\times$ 10$^{-5}$   & 1.4 $\times$ 10$^{-5}$     \\ 
N {/} H &  6.4 $\times$ 10$^{-5}$   &  2.1 $\times$ 10$^{-5}$       \\ 
O {/} H &   6  $\times$ 10$^{-5}$   &   2.0 $\times$ 10$^{-5}$    \\ 
S {/} H  &  8.0  $\times$ 10$^{-8}$  &  8.0   $\times$ 10$^{-8}$     \\
Fe / H & 1.5  $\times$ 10$^{-9}$  &  1.5   $\times$ 10$^{-9}$     \\ 
  \hline                    
  \hline                  
\end{tabular}
\end{table}
We consider two different models as displayed in Table \ref{tab:model}. Model (a) may be considered as a template of TMC1,  and assumes a  density of hydrogen nuclei  n$_H$ = 2 $\times$ 10$^4$ cm$^{-3}$.   The elemental abundance of carbon relative to hydrogen nuclei is taken as 4.15 $\times$ 10$^{-5}$  to reproduce the derived relative abundance of CO \citep{ohishi:92}. We derive the oxygen elemental abundance by imposing a C/O ratio of 0.7 appropriate for TMC1 and take the nitrogen elemental abundance  used by \cite{legal:14} in their work on nitrogen chemistry. The elemental abundance of sulfur is not well constrained and we have taken the low metal case value of 8.0   $\times$ 10$^{-8}$.
Model (b) is  more representative of  a pre stellar core with a density of 2 $\times$ 10$^5$ cm$^{-3}$  similar to L134N or Barnard 1 (B1) where the elemental abundances of carbon, oxygen and nitrogen are reduced by a factor  of 3 % and nitrogen by a factor of 3 
to account for depletion. %The C/O ratio is assumed to be the same and equal to 0.7. 
 T~=~10K  in both cases and the cosmic ionization rate $\zeta$ per H$_2$ is 1.3  $\times$ 10$^{-17}$ \textbf{s$^{-1}$}  as in \cite{legal:14}.
\subsection{Results}

We summarize  our results  obtained with a 10$^{-3}$ value of the o/p ratio of  {\HH}  in Table \ref{tab:res} and give some observational values for comparison. Time dependent effects may be  visualized
 from the values reported at 10$^6$ years and at steady state for model (a). 
 Steady state is reached after a few 10$^7$ and 10$^6$ years respectively for models (a) and (b).
 As there are  fewer $^{15}$N enrichment reactions  than previously assumed \citep{terzieva:00} %due to the small values of the energy transfer,
 most nitrogen containing species are found to have isotopic abundance ratios close to the solar value ($^{14}$N/$^{15}$N = 440) 
given by the Genesis mission \citep{marty:11}. 
\begin{table*}[h]
\caption{Model (a) and (b) results and observations. ss means stationary state.  The value of the o/p ratio of {\HH} is taken as 10$^{-3}$.  }   
\label{tab:res}      
\centering          
\begin{tabular}{l|cc|c|lll}     % 3columns 
\hline\hline       
     & \multicolumn{2}{c|}{Model (a)} &    Model (b) & TMC1 & L1544 & B1\\
     & t= 10$^6$ yrs& ss & ss &   &    &    \\  
 %   &  dcn_nom_A5a  &   dcn_nom_A5b     
     \hline
%density n$_H$ (cm$^{-3}$) &    2 $\times$ 10$^4$ &   2 $\times$ 10$^5$  \\
electronic fraction & 1.4 $\times$ 10$^{-8}$   &   2.8 $\times$ 10$^{-8}$   &   1.7 $\times$ 10$^{-8}$ &   & &   \\
N / \fN  & 440 & 456  & 455 &     &  &    \\   
2 $\times$ N$_2$ / \fN N &   438 &431& 437& &  & \\
NH / \fNH  & 429 &  428  &  421 &     &  &    \\  
NH / ND  & 16  &  31  &   9 &     &  &    \\   
\NHHH / \HH    &  6.7 10$^{-9}$  &      1.3 10$^{-9}$   & 6.0 10$^{-9}$& 2 10$^{-8}$   $^{(8)}$ &   &  \\
\NHHH / \fNHHH  &  333 &      386       &      387  & &   & 300$^{+ 55}_{-40}$ $^{(1)}$\\
\NHHD / \HH    &   3.8 10$^{-10}$ &     5.8 10$^{-11}$        &  3.3 10$^{-10}$  & 4 10$^{-10}$  $^{(9)}$&   &  \\
\NHHD / \fNHHD  &  215 &      276  &      336  &  &  & 230$^{+ 105}_{-55}$ $^{(1)}$ \\
\NHHH / \NHHD  &  18  &     22          &     18  & 50 $^{(9)}$&  &  \\
\NNHp / \HH    &   4.8  10$^{-10}$ &     1.3  10$^{-10}$         & 2.1  10$^{-10}$ & 5 10$^{-10}$  $^{(8)}$ &   &  \\
\NNHp / \NfNHp &  431 &  430   &   423  &  &1050$^{\pm  220}$ $^{(2)}$  &   400$^{+ 100}_{-65}$ $^{(1)}$\\
\NNHp / \fNNHp & 437 &   432  &   433   &  &   1110$^{\pm  240}$ $^{(2)}$  &   $>$ 600 $^{(1)}$  \\
\NNHp / \NNDp  &  16 &  29  &  8.6 & 12.5 $^{(9)}$   & & 2.9 $^{(1)}$  \\
CN / \HH    &   6.8 10$^{-9}$ &      5.5  10$^{-9}$  &  1.2  10$^{-9}$  & 3 10$^{-8}$ $^{(8)}$ &   &  \\
CN / \tCN   & 67 &   84  &    63    &   & &50$^{+19}_{-11}$ $^{(1)}$\\
CN / \CfiN  &  430 &  449  &   445 &   & & 240$^{+135}_{-65}$ $^{(1)}$\\
\tCN / \CfiN  & 6.4 &   5.3  &  7.0 &   & 7.5$^{\pm 1}$ $^{(3)}$  &\\
HCN / \HH    &   7.4  10$^{-9}$&   5.9  10$^{-10}$ &  5.4  10$^{-10}$ & 2 10$^{-8}$  $^{(8)}$ &   &  \\
HCN / \tHCN  & 93 &  168 &   114  &   &  & 30$^{+7}_{-4}$ $^{(1)}$\\
HCN / \fHCN  &  398 & 445  &    453   &  &   &  165$^{+30}_{-20}$ $^{(1)}$\\
 \tHCN /  \fHCN & 4.3 & 2.6  &  4.0  &  2 - 4.5 $^{(5)}$  &   & 5.5 $\pm$1 $^{(1)}$ \\
HCN / DCN  &  43 &  96 &  22 &  &   &  20$^{+6}_{-10}$ $^{(1)}$\\
HNC / \HH    &   5.6 10$^{-9}$ &    7.4  10$^{-10}$  &   8.4  10$^{-10}$   & 2 10$^{-8}$  $^{(8)}$ &   &  \\
HNC/ \tHNC &  93 & 180   & 121 & 54 - 72 $^{(4)}$ &   & 20$^{+5}_{-4}$ $^{(1)}$\\
HNC/ \fHNC &  405 & 442   & 446 & 250 - 330  $^{(4)}$  &   &  75$^{+25}_{-15}$ $^{(1)}$ \\
{\tHNC} /  {\fHNC} &  2.5 & 1.75   & 3.7   &  4.6 $\pm$ 0.6  $^{(4)}$ &   & 3.7 $\pm$ 1 $^{(1)}$ \\
HNC/ DNC &  23 & 66   &  16 &  &   &  2.9$^{+1.1}_{-0.9}$ $^{(1)}$\\
%DNC/ DN\tC &  5.1 & 12   &  13 &  &   &  30$^{+8}_{-5}$ (1) \\
NO / \HH    &  1. 10$^{-7}$  &  3.1  10$^{-8}$  & 4.1  10$^{-8}$ & 2.7 10$^{-8}$  $^{(10)}$ &   &  \\
NO / {\fN}O  & 438 & 451  &  446  &     &  &    \\ 
CO / \HH    &  8.1 10$^{-5}$ &    8.0 10$^{-5}$ &  2.8  10$^{-5}$       & 8 10$^{-5}$  $^{(8)}$ &   &  \\
  {CO} /  {\tC}O & 68 & 67.4 &  68 &    &   &     \\
CH / \HH    &   1.3 10$^{-9}$  & 1.7 10$^{-9}$&  1.7  10$^{-10}$  & 2 10$^{-8}$ $^{(8)}$  &   &  \\
{CH} /  {\tC}H & 74 & 154  &  71 & $>$ 71 $^{(6)}$   &   &     \\
\HCOp / \HH    &  2.5 10$^{-9}$  &    3.610$^{-10}$    & 5.7  10$^{-10}$ & 8 10$^{-9}$  $^{(8)}$ &   &  \\
{\HCOp} / {\HtCOp} & 56 & 65 & 56 &    &   & 59 $^{(7)}$  \\
{\HCOp} / {\DCOp} & 15 & 29  & 8.4  &  50 $^{(9)}$  &   &     \\
\hline                    
  \hline                  
\end{tabular}
\tablebib{(1)~\citet{daniel:13};
(2) \citet{bizzocchi:13}; (3) \citet{hilyblant:13}; (4) \citet{liszt:12}
; (5) \cite{hilyblant:13b}; (6) \citet{sakai:13b}, (7) \citet{hirano:14} assuming CO/ C$^{18}$O=500,
(8) \cite{ohishi:92}, (9) \cite{tine:00}, (10) \cite{gerin:93}.
}
\end{table*}
\section{Discussion}
\begin{figure*}%[h]
   \centering
  \includegraphics[width=14cm]{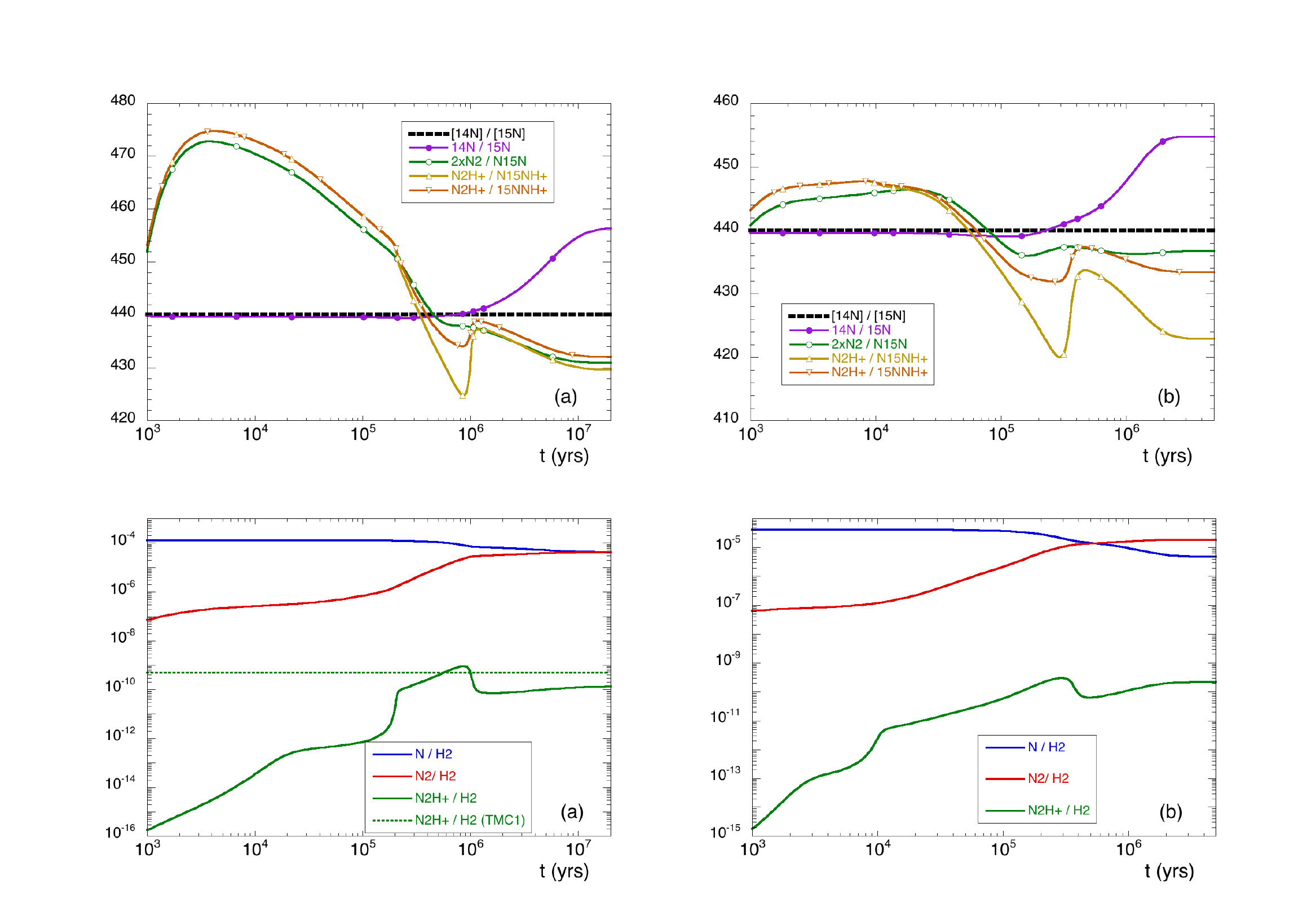}
   \caption{ Upper panel : time dependence of N / {\fN} isotopic ratios in atomic and molecular nitrogen  and {\NNHp} ions. (a) and (b) correspond to the models defined in Table \ref{tab:model}; the black heavy dotted line represents the elemental N / \fN.  Lower panel :  time dependence of the fractional abundances relative to {\HH} of N, {\NN} and {\NNHp} for models (a) and (b) . The value reported for {\NNHp} towards TMC1 \citep{ohishi:92} is displayed as a horizontal green dashed line.}
              \label{fig:nn2}%
    \end{figure*}

We  display the time dependence of various isotopic ratios  and  fractional abundances relative to {\HH}  and discuss the chemical behavior involved 
in the fractionation processes for the two reported models. We first %show 
 consider in Figure \ref{fig:nn2}  the reservoirs of nitrogen, atomic and molecular nitrogen as well as {\NNHp}  ions which are chemically linked to \NN.
Atomic nitrogen becomes depleted in {\fN} after about 10$^5$ years whereas molecular nitrogen is slightly enriched. 
 These evolution times are also required to build significant amounts of molecular compounds which compare satisfactorily with available observations. 
 The overall dependence of {\NNHp} follows closely that of {\NN} as it is formed from {\NN} + {\HHHp} reaction, with a slight decoupling between {\fNNHp}  and {\NfNHp} at long evolution times. 
We find that the isotopic ratio of the   {\NNHp} ions displays an almost constant value close to the solar value after some 10$^5$ years. They are in good agreement with observations in B1 but disagree by a factor of 2 for L1544.
 The trend that {\fNNHp} is less abundant than {\NfNHp} is reproduced in our results, as a result of the differences of endothermicity in their reactions with \NN.  We checked that introducing \fN$_2$ and species containing two {\fN} atoms had no effect on these ratios.
We could not find any gas-phase mechanism able to generate 
 such a large ratio in pre-stellar core conditions. The high isotopic ratio found in L1544 implies an equivalently large ratio for molecular nitrogen, which is in strong contradiction with our findings.
%showing almost no enrichment. This results is due to the fact that the NNH$^{+}$/$^{15}$NNH$^{+}$/N$^{15}$NH$^{+}$ abundances are controlled by the N$_2$/$^{15}$NN + H$_3$$^{+}$ 
%and NNH$^{+}$/$^{15}$NNH$^{+}$/N$^{15}$NH$^{+}$ + e$^{-}$ reactions, the fractionation reaction, $^{15}$NN + NNH$^{+}$, involving too small fluxes to play an important role. 
These results are markedly different from those derived by \cite{hilyblant:13} who found a moderate {\fN} enrichment as these authors introduced the {\fN} + {\NNHp}  fractionation reaction  which we have found not to occur. \footnote{These authors also interchanged the endothermicities of the {\fNNHp} and {\NfNHp} reactions with N and {\NN}.}%,  which was discarded in our chemical network as discussed in Section \ref{sec:reac}}.
\subsection{Nitrogen hydrides}
\label{sec:NH}
\begin{figure*}%[h]
   \centering
  \includegraphics[width=14cm]{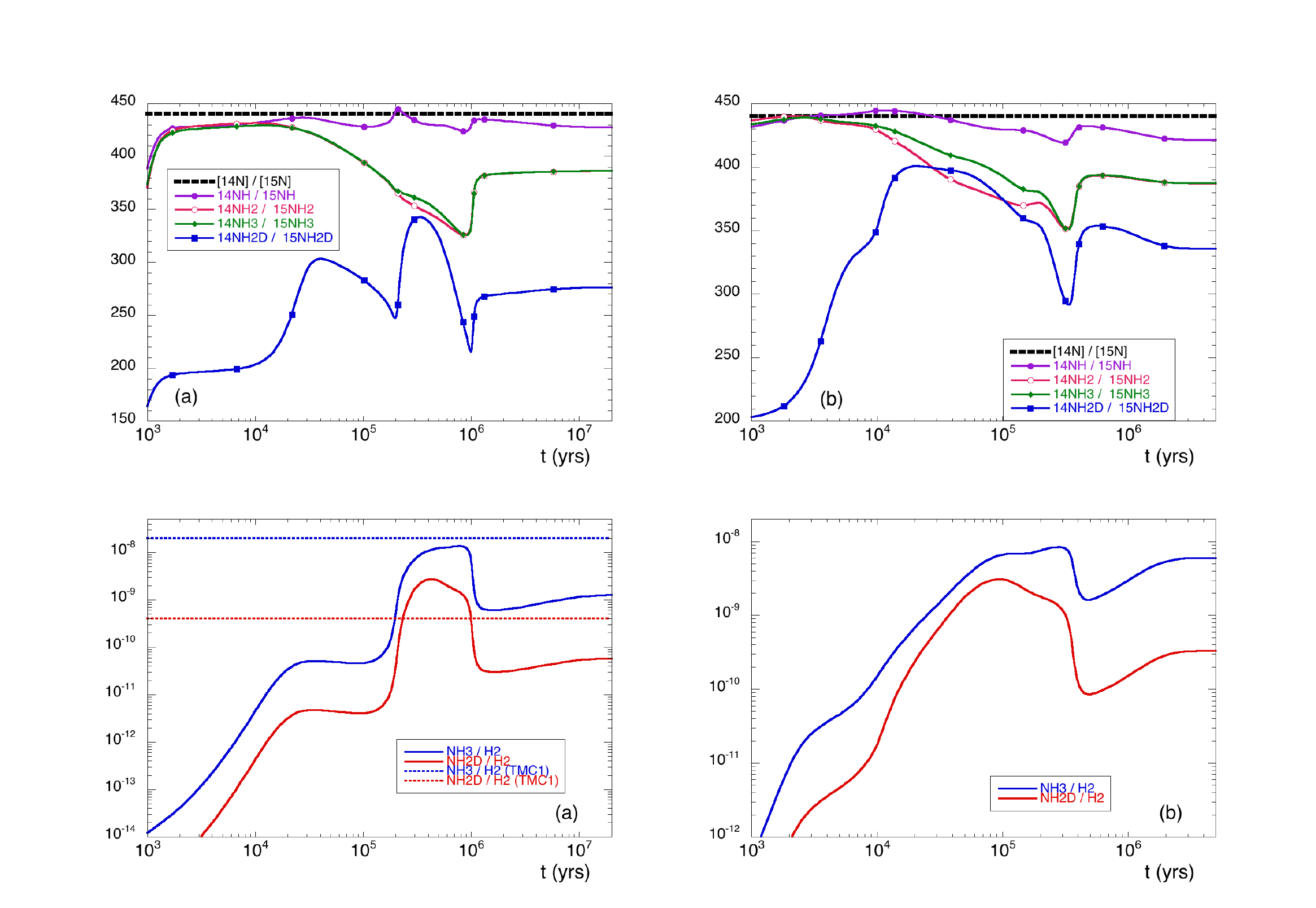}
   \caption{ Upper panel :   Time dependence of N / {\fN} isotopic ratios in nitrogen hydrides. (a) and (b) correspond to the models defined in Table \ref{tab:model}; The black heavy dotted line represents the elemental N / \fN.
   Lower panel :  time dependence of the fractional abundances relative to {\HH} of NH, {\NHH}, {\NHHH} and {\NHHD} for models (a) and (b). The values reported for {\NHHH} \citep{ohishi:92} and {\NHHD} \citep{tine:00} towards TMC1 are displayed as  horizontal blue and red dashed lines respectively in the left panel.}              \label{fig:nh3-fN}%
    \end{figure*}

We display in Figure \ref{fig:nh3-fN} the time evolution of the isotopic ratios of  nitrogen hydrides   and of their fractional abundances relative to \HH. {\NHHH} and {\NHH} have a very similar behavior as they both result from the reaction chain  starting with the {\Np} + {\HH} reactions. 
 The species NH$_2$ and NH$_3$ are found to be enriched in $^{15}$N due to the $^{15}$N$^{+}$ + o-H$_{2}$ reaction which has a slightly smaller (weak) endothermicity, as reported in Table \ref{tab:npHH},    than the 
corresponding $^{14}$N$^{+}$ + o-H$_2$ reaction, a difference which  slightly favors  $^{15}$NH$^{+}$ formation. Nevertheless, despite the large uncertainties regarding 
the rate of the N$^{+}$ + H$_2$ reaction, our results are in fair agreement with the results of \cite{daniel:13} for prestellar core B1. % for $^{14}$NH$_3$ /$^{15}$NH$_3$ 
%(300$^{+ 55}_{-40}$). 
As the N$^{+}$ + HD $\rightarrow$ ND$^{+}$ + H reaction has a smaller endothermicity than the N$^{+}$ + H$_{2}$ $\rightarrow$ NH$^{+}$ + H reaction, 
the modeled $^{14}$NH$_2$D / $^{15}$NH$_2$D ratio behaves somewhat differently than the $^{14}$NH$_3$ / $^{15}$NH$_3$  ratio.  The ratio exhibits large variations around 10$^5$ years and becomes smaller than that of  {\NHHH}  at large times and at steady state. %with relatively large variations as a function of 
%initial conditions that may explain the various observations \cite{daniel:13,gerin:09}, all measurements being 
The observed values are compatible with calculations at sufficiently large times. The values of this ratio reported in \cite{gerin:09} have been found to be too large as a result of assuming a single excitation rotational temperature. The future availability  of collisional excitation rates of {\NHHD} by {\HH} \citep{daniel:14} may give rise to additional changes.
 NH  does not follow the same trend as  NH$_2$ and NH$_3$ and is  only  slightly enriched in \fN.%, as its formation results principally
  As discussed by \cite{hilyblant:13}, NH is mainly formed 
 from the dissociative recombination (DR)  of \NNHp,  a reaction which has been recently revisited by \cite{vigren:12}
 who derive a branching ratio towards NH of 7\%.
 We have checked that this analysis still holds for model (b) even if  the  reaction of  {\NNHp} with CO becomes more efficient.
The formation route of NH through NH$_2^+$ recombination may take over for highly depleted CO. 
%   noticed by \cite{hilyblant:13}.  %does not follow the same trend as NH$_2$  and NH$_3$.
%N$^{15}$NH$^{+}$/NNH$^{+}$ = 446 $\pm$ 71 in molecular cloud L1544 
%\cite{bizzocchi:10}, as well as with the ratio  {\NfNHp} / {\NNHp} = 400$^{+ 100}_{- 65}$ detected in the  B1 prestellar core \citep{daniel:13}. 
%It should be noted however that our calculations are not in agreement with their respective measurements for 
%{\fNNHp} / {\NNHp} = 1000 $\pm$ 200 \citep{bizzocchi:13} 
%and {\fNNHp} /  {\NNHp}  > 600 \citep{daniel:13} as we find similar isotopic ratios for {\NfNHp} /  {\NNHp}  and  {\fNNHp} /  {\NNHp}  in our calculations. However, 

\subsection{Nitriles and isonitriles}
Deriving {\fN} isotopic ratios of CN, HCN and HNC from observations is a difficult challenge as the transitions of the main isotopologues are optically thick. %A comparison with observations for CN, HCN and HNC  is not obvious as the lines of the main isotopologues of these species are optically thick. 
Then, most of the reported observational values of the {\nN} / {\fN} molecular ratios 
 are obtained from the ratios of the minor isotopologues $^{13}$CN / C$^{15}$N, H$^{13}$CN / HC$^{15}$N  
 and HN$^{13}$C / H$^{15}$NC which is subsequently multiplied by an assumed C / {\tC}  value, usually taken as 68 \citep{milam:05}. 

\subsubsection{ {\tC}/ {\fN} ratios}
 We test these hypotheses in our models by explicitly introducing fractionation reactions of \tC, as discussed in Section \ref{sec:reac}.
 Table \ref{tab:res} shows that
% that this assumption  holds well for CO, which is the reservoir of C in those conditions, but 
the {\tC} isotopic ratios of CN, HCN and HNC vary both with time and density. 
      \begin{figure*}%[h]
   \centering
  \includegraphics[width=14cm]{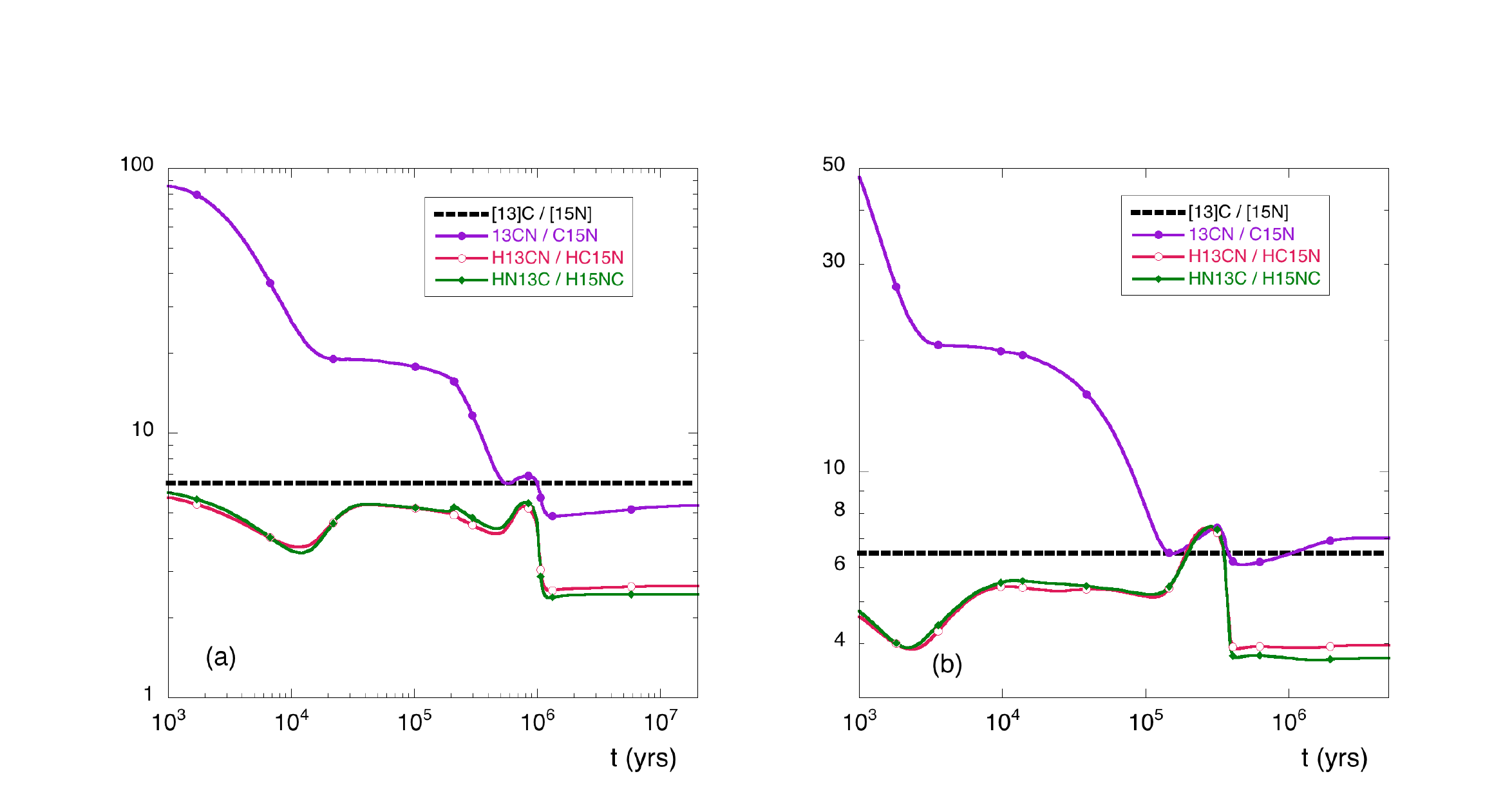}

   \caption{Time evolution of \tC / {\fN} ratios in CN, HCN and HNC for models (a) and (b). The \tC/{\fN} elemental ratio is displayed as heavy dotted line.}
              \label{fig:1315}%
    \end{figure*}
Figure \ref{fig:1315} displays the  $^{13}$CN / C$^{15}$N, H$^{13}$CN / HC$^{15}$N  
 and HN$^{13}$C / H$^{15}$NC ratios as a function of time for the two considered models.
 The deviation from the elemental ratio of 6.48 is %notably different 
 significant  for HCN, HNC and CN.    % than for CN.
 %There is a narrow range of times where the hypothesis (\tC / {\fN}  ratio in CN, HCN and HNC equal to 6.48)  is fulfilled approximately in both models. %With the lower density conditions and larger amounts of gas phase C and N corresponding to model (a), the deviations from 6.48 of HCN, HNC and CN  tend to increase at steady state. 
%The steady state values of the \tC/\fN ratio  are   slightly  larger than the elemental value for CN and significantly lower than the elemental value for HCN and HNC. 
 %
\subsubsection{{\tC} chemistry}
\begin{figure*}%[h]
   \centering
  \includegraphics[width=14cm]{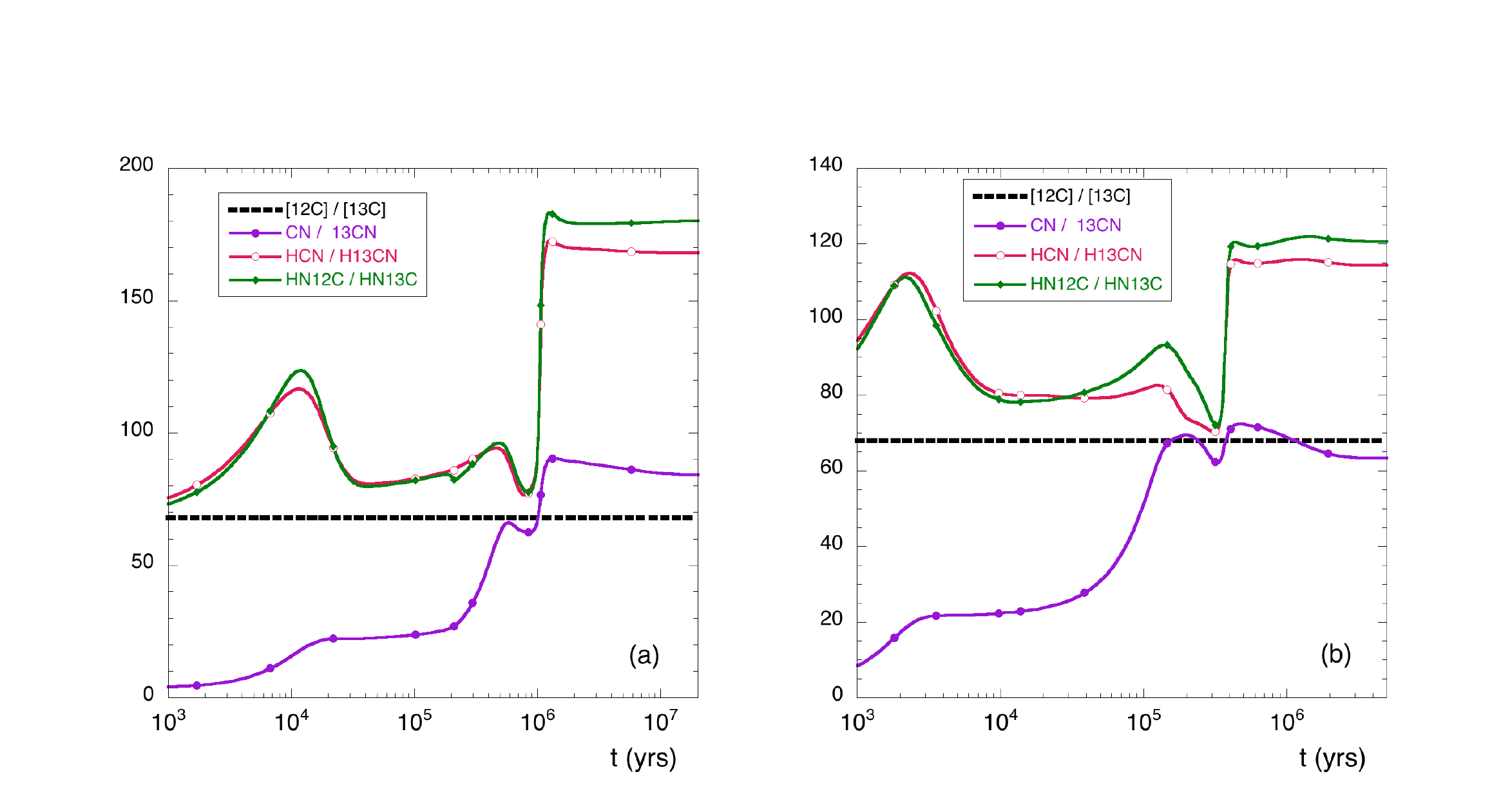}
   \caption{time dependence of C / {\tC} isotopic ratios in CN, HCN and HNC. (a) and (b) correspond to the models defined in Table \ref{tab:model}}
              \label{fig:tCN}%
    \end{figure*}

\begin{figure*}%[h]
   \centering
  \includegraphics[width=14cm]{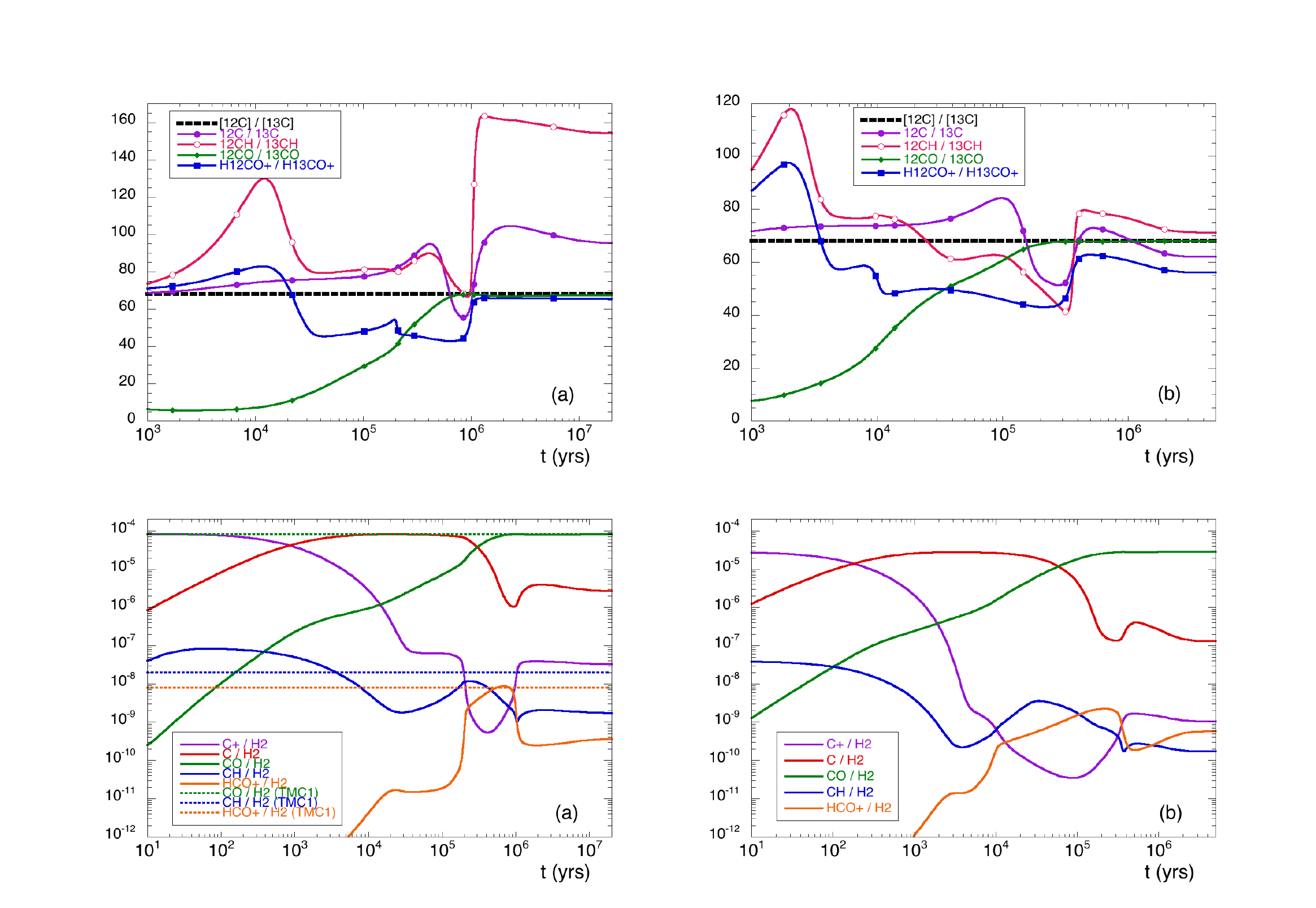}
   \caption{ Upper panel : Time dependence of C / {\tC} isotopic ratios in C, CH, CO and HCO$^+$. The black heavy dotted line represents the elemental {\dC} / \tC.
   Lower panel :  Time dependence of the fractional abundances relative to {\HH} of C,  \Cp, CO, CH and {\HCOp},   for models (a) and (b). The observational values  towards TMC1 \citep{ohishi:92}  are displayed as  horizontal   dashed lines  with corresponding colors in the left panel. (a) and (b) correspond to the models defined in Table \ref{tab:model}}
              \label{fig:tC}%
    \end{figure*}

We now consider the time dependences  of   the $^{12}$C / $^{13}$C isotopic ratios in   CN, HCN and HNC  species as displayed in Figure \ref{fig:tCN}. %exhibit a complex behavior (except for CO)
The time dependent ratio displays large variations, which is
%In contrast to nitrogen, the $^{12}$C/$^{13}$C ratio for these molecules shows a complex behavior
 due to the fact that there are various reactions incorporating $^{13}$C 
in the  molecules. %Consequently, the use of $^{13}$C containing minor isotopologues as a proxy for the $^{12}$C ones %(taking into consideration the $^{12}$C/$^{13}$C ratio equal to 68) 
%is questionable. %could lead to substantial errors in the calculated abundances of the major isotopologues. 
%As carbon chemistry is  significantly much more complex than that of nitrogen  in
%molecular clouds, we describe in detail our results for $^{13}$C enrichment before comparing observations and calculations for HCN, HNC and CN.
The elemental value of the ratio (68) is fulfilled in a narrow range around 1 Myr for model (a) and 2 $\times$ 10$^5$ yrs for model (b)  but steady state values are significantly different  except for CN.  
This relatively complex behavior results from the many different reaction channels involved in {\tC} chemistry. 
We then also consider other {\tC} containing species and display  the {\dC}  / {\tC}  ratio in C, CH, CO and HCO$^+$
in Figure \ref{fig:tC}  as well as their fractional abundances relative to \HH.
The $^{12}$C/$^{13}$C isotopic ratios of the various molecules are highly dependent %on the initial conditions and 
on the evolution time. 
The transition from gas-phase atomic carbon  toward CO controls the $^{13}$C enrichment. As long as there is still a relatively 
high carbon atom concentration in the gas phase, there is enough free $^{13}$C  to allow strong enrichment of CN
through the {\tC} + CN reaction. When  CO molecules become the reservoir of carbon,  even if in that case the $^{13}$C concentration is low, the $^{13}$C$^{+}$ + $^{12}$CO  $\rightarrow$ 
$^{12}$C$^{+}$ + $^{13}$CO reaction still leads to a small  $^{13}$CO enrichment \citep{langer:90,milam:05}. Although this  small excess is not measurable in CO, significant amounts of $^{13}$C are locked up in CO and  most of the other carbon containing species become depleted in {\tC}, as found for CH and other carbon chains. 
 This effect is seen in Figure \ref{fig:tC} for the isotopic ratio of CH in model (a) at steady state where CO is slightly enriched,
leading to a significant depletion of {\tC} in CH. 
Our results 
for C and {\HCOp} are similar to those of  \cite{furuya:11} who studied  the {\tC} fractionation of multiple carbon chains by 
explicitly introducing the dependence of the {\tC} position in the chain.
%{\bf{We have introduced the proton transfer reaction between {\HCOp} and CN which is   exothermic
%and have checked, at the DFT level, the absence of any barrier.}}
We see that %both CN and 
{\HCOp} is marginally enriched in {\tC} at steady state whereas HCN and HNC are significantly depleted in \tC. 
 We also note that CN and {\HCOp} may react via proton transfer  to give CO + HNC$^+$ and have checked
, at the DFT level, the absence of any barrier. The corresponding reaction rate is 2.2 $\times$ 10$^{-9}$ ($\frac{T}{300}$)$^{-0.4}$
 cm$^{3}$ s$^{-1}$ when using the capture rate theory. This reaction has not been included in any chemical database up to now. 
%The isotopic variants of this reaction couple the {\tC} and  {\fN}  chemistries.
 
HNC has been shown recently to react with atomic carbon \citep{loison:14}, which leads to the different steady state isotopic ratios obtained for HCN and HNC. 
 The CN chemistry is then  somewhat decoupled from that of   HCN and HNC.   HCN and HNC are formed at relatively long times via CN + H$_3$$^{+}$ $\rightarrow$ HNC$^{+}$ /HCN$^{+}$ + H$_2$, followed immediately by HNC$^{+}$ + H$_2$ $\rightarrow$ HCNH$^{+}$ + H giving back HCN and HNC via DR \citep{mendes:12}. With the adopted elemental abundances, the main CN destruction reactions are however O + CN and N + CN so that the HCNH$^{+}$/HCN$^{+}$/HNC$^{+}$/HCN/HNC/CN 
 network is not a closed system in contrast to models including coupled gas-grain chemistry \citep{loison:14}.

 \subsubsection{ {\fN} fractionation} 
\begin{figure*}%[h]
   \centering
  \includegraphics[width=14cm]{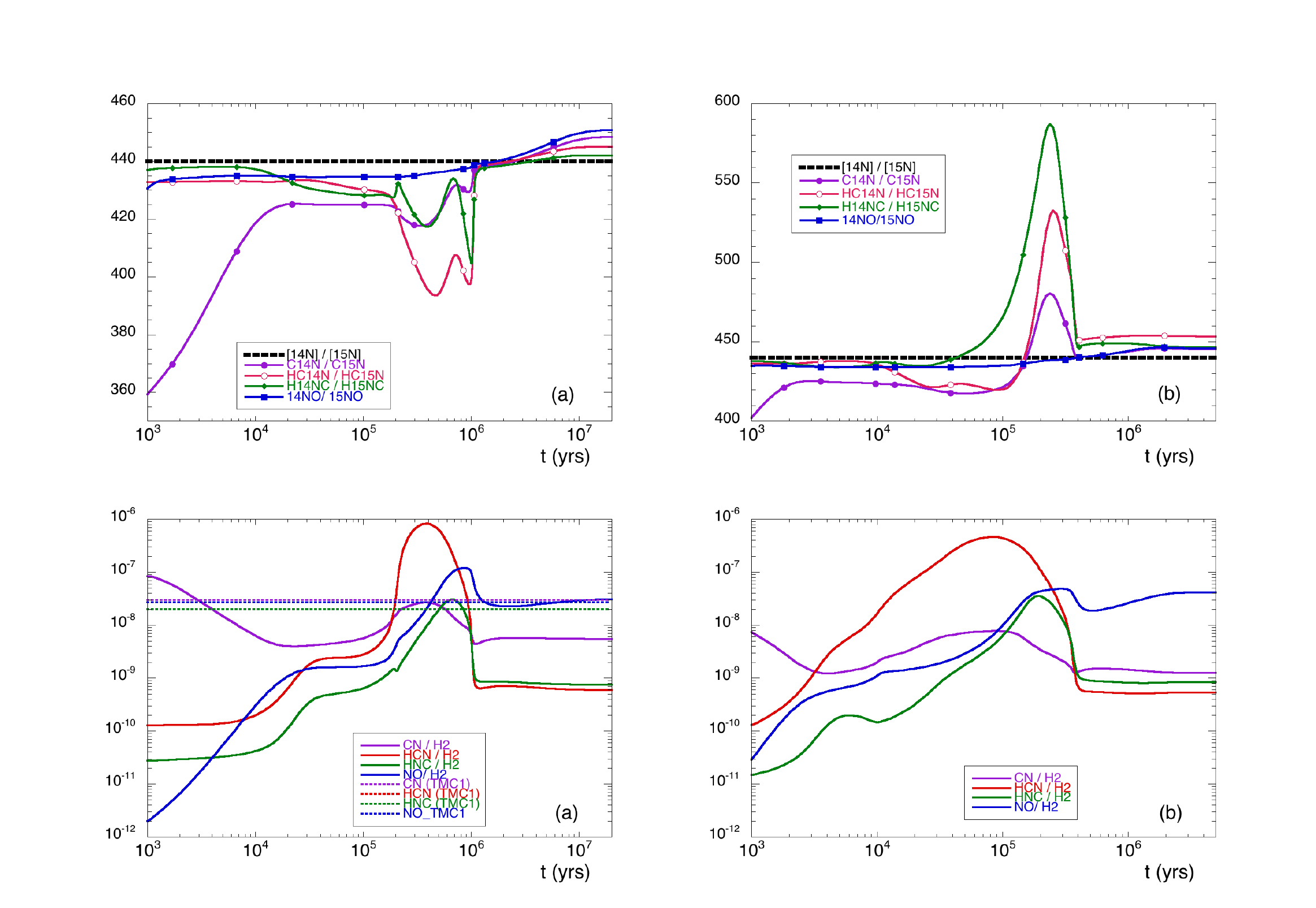}
   \caption{Time dependence of N / {\fN} isotopic ratios in CN, HCN, HNC and NO. (a) and (b) correspond to the models defined in Table \ref{tab:model}}
     \label{fig:15CN}%
    \end{figure*}

\begin{figure*}%[h]
   \centering
  \includegraphics[width=14cm]{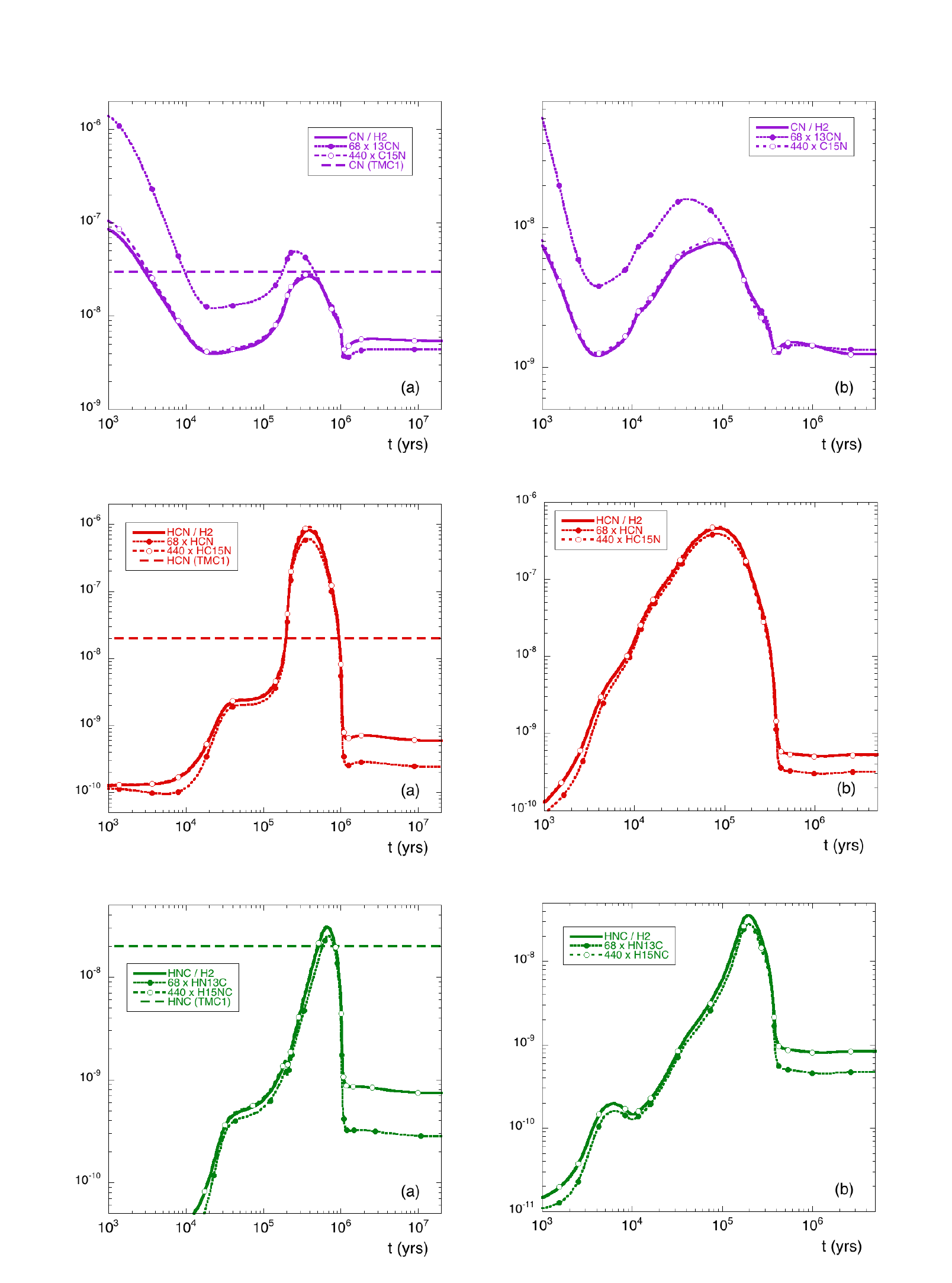}
   \caption{Time dependence of   CN, HCN, HNC and their isotopologues. (a) and (b) correspond to the models defined in Table \ref{tab:model}}
      \label{fig:td}%
    \end{figure*}

We now display also the N / {\fN} fractionation in nitriles as well as that of NO in Figure~\ref{fig:15CN}.
The time dependences of the isotopic ratios are markedly different in the  two models, except for NO. Whereas CN, HCN and HNC are somewhat enriched in {\fN} for model (a) both at intermediates times and steady state, the opposite result is obtained in model (b)  after 10$^{5}$ years.  We display as well the time dependences of the fractional abundances of these molecules in Figure \ref{fig:td} in order to better understand the previously described differences observed in the various fractionation ratios. 
%
%However, we find that CH and C2 are significantly depleted in \tC, as well as HCN and HNC.

Our model (a) is intended to be representative of TMC1, a moderately dense cloud in an early evolutionary stage.  
% The chemical evolution time is estimated to be between 2$\times$10$^{5}$ and 1$\times$ 10$^{6}$ years from the comparison between  observations and models for  most molecular compounds. This range leads to a variation in the $^{12}$C/$^{13}$C ratio between 70 and 100 for H$^{13}$CN and HN$^{13}$C, and a value varying 
%from 20 at 2$\times$10$^{5}$ yrs to 70 at 1$\times$10$^{6}$ yrs for $^{12}$CN/$^{13}$CN. 
We see that the various abundances and ratios  are very sensitive to the chosen "age", assumed to be the relevant chemical evolution time. Considering a  TMC1 age of 1Myr  leads to a reasonable agreement with the sparse  observations available, see Table \ref{tab:res}.  
Model (b) is more likely to be representative of a denser  evolved molecular cloud such as L134N, L1544, Barnard B1, ... where elemental C, O, N are partially depleted through sticking  on grains.  Steady state values obtained with significant depletion conditions \citep{roueff:05} may be used as corresponding proxies. We see that ammonia isotopologues are satisfactorily accounted for in our model. The agreement between observed and calculated \tC / {\fN} ratios is somewhat misleading as the modeled value mainly results from a significant depletion in the {\tC} species. The ratios involving N / {\fN}  are found  close to the elemental value in our model (b) at steady state  (although {\NHH} and {\NHHH} are somewhat enriched in {\fN} through the {\fNp} + {\HH} reaction as explained in subsection \ref{sec:NH}), which is in agreement with the fact that  no significant 
gas phase fractionation mechanisms have been  found. The occurrence of small ratios in B1 observations \citep{daniel:13},
if real, implies other mechanisms at work. An obvious suggestion lies in the processes involved in adsorption/{\bf{desorption}}   on grains and possible surface reactions.
\subsection{Role of the {\HH} ortho to para ratio}
We now test the role of the value of the ortho/para ratio of molecular hydrogen (OPR) in the context of ammonia chemistry and fractionation determination. We run two additional models in the frame of models (a) and (b) by changing the OPR by a factor of 10 
both upwards and downwards. The abundances of p-{\HH} and o-{\HH} are expressed respectively as
n(p-{\HH}) = $\frac{1}{1+\rm{OPR}}$ n(\HH) and 
n(o-{\HH})~=$\frac{\rm{OPR}}{1+\rm{OPR}}$ n(\HH).
This ratio is introduced, in addition to the reaction of  {\HH} with  {\Np} (\fNp), in the reverse
reaction of the fractionation of \HHHp, namely the \HHDp + {\HH} reaction which  plays a significant role in the deuterium 
fractionation of various molecules \citep{pagani:11} as shown below:
\\
\\
\begin{tabular}{llll}
{\bf{$\HHDp  + \rm{p}-\HH$}} &  $\rightarrow$  & {\bf{$ \HHHp  + \HD$}}& , k$_1$  (cm$^3$ s$^{-1}$)\\
{\bf{$\HHDp  + \rm{o}-\HH$}}  & $\rightarrow$ &   {\bf{$ \HHHp  + \HD$}}& , k$_2$   (cm$^3$ s$^{-1}$)\\\  
\end{tabular}
\\
with k$_1$ = 2.0 $\times$ 10$^{-9}$ $\times$ exp(-232/T)  and 
k$_2$ =  2.0 $\times$ 10$^{-9}$ $\times$ exp(-61.5/T) . 
\begin{table*}[h]
\caption{Dependence of the fractionation ratios on the o/p ratio of {\HH} for model (a). ss means stationary state.    }   
\label{tab:opa}      
\centering          
\begin{tabular}{|l|cc|cc|cc|}     % 3columns 
\hline\hline       
     & \multicolumn{2}{c|}{OPR =  10$^{-4}$} &    \multicolumn{2}{c|}{OPR =  10$^{-3}$} &\multicolumn{2}{c|}{OPR =  10$^{-2}$}\\
     & t= 10$^6$ yrs& ss &  t= 10$^6$ yrs& ss  &    t= 10$^6$ yrs& ss    \\  
 %   &  dcn_nom_A5a  &   dcn_nom_A5b     
     \hline
%density n$_H$ (cm$^{-3}$) &    2 $\times$ 10$^4$ &   2 $\times$ 10$^5$  \\
electronic fraction & 5.1 $\times$ 10$^{-8}$  &  2.2 $\times$ 10$^{-7} $   &4.8 $\times$ 10$^{-8}$   &   2.1 $\times$ 10$^{-7}$   &  3.7 $\times$ 10$^{-8}$ &  4.2 $\times$ 10$^{-8}$    \\
N / \fN  & 440 &   456  & 440 & 456 &  440   & 452   \\   
2 $\times$ N$_2$ / \fN N &  438&   430  &  438 &431&  438  &  437  \\
NH / \fNH  &  431 &  429   & 429 &  428  &  426  &  418  \\  
NH / ND  &  16  &   31  & 16  &  31  &   18   &  23 \\   
\NHHH / \fNHHH  & 345 &  409   &   333 &       386          &    374   &  395   \\
\NHHD / \fNHHD  &  206 &  267  &  215 &      276         &     265  &  303  \\
\NHHH / \NHHD  & 8   &    14 & 17  &      22        &   43  &  63 \\
\NNHp / \NfNHp &431  &  429   & 431 &  430  &    429 &  421\\
\NNHp / \fNNHp & 437 &  432   & 437 &   432  &  436  &  433  \\
\NNHp / \NNDp  & 16 &   30  & 16 &  29  & 17 &  21  \\
CN / \tCN   &  67 &  85   & 67 &   84  &  67   &  70 \\
CN / \CfiN  & 432 &   449  & 430 &  449  &  429 & 434    \\
HCN / \tHCN  & 92 & 168   &   93 &  168 &    93   &  165 \\
HCN / \fHCN  & 401  & 453    &  398 & 445  &   400    &  413 \\
HCN / DCN  &  44 &   95  & 43 &  96 &  40  &  55  \\
HNC/ \tHNC &  91 &   178  & 93 & 180   &  97   &  195   \\
HNC/ \fHNC & 410 &   451   & 405 & 442   &  405 &   410   \\
HNC/ DNC &  23 &  66  & 23 & 66   &  22   &  32  \\
%DNC/ DN\tC &  5.1 & 12   &  13 &  &   &  30$^{+8}_{-5}$ (1) \\
 % {CO} /  {\tC}O & &     & 67 & 68 &      &       \\
%{CH} /  {\tC}H & &     & 78 & 101  &      &       \\
{\HCOp} / {\HtCOp} & 57&  66   & 56 & 65 &   54  &  55   \\
{\HCOp} / {\DCOp} &  15 &  30  & 15 & 29  &   16 &  19   \\
\hline                    
  \hline                  
\end{tabular}
\end{table*}
 \begin{table}[h]
\caption{Dependence of the fractionation ratios on the o/p ratio of {\HH} for model (b) at steady state.    }   
\label{tab:opb}      
\centering          
\begin{tabular}{l|c|c|c}     % 3columns 
\hline\hline       
     &  {OPR =  10$^{-4}$} &     {OPR =  10$^{-3}$} & {OPR =  10$^{-2}$}\\
%     & t= 10$^6$ yrs& ss &  t= 10$^6$ yrs& ss  &    t= 10$^6$ yrs& ss    \\  
 %   &  dcn_nom_A5a  &   dcn_nom_A5b     
     \hline
%density n$_H$ (cm$^{-3}$) &    2 $\times$ 10$^4$ &   2 $\times$ 10$^5$  \\
electronic fraction & 2.2$\times$ 10$^{-8}$  &  1.7 $\times$ 10$^{-8} $   &1.1 $\times$ 10$^{-8}$      \\
N / \fN  & 457 &   455  & 445  \\   
2 $\times$ N$_2$ / \fN N &  436&   437  & 438  \\
NH / \fNH  &  425 &  421   & 418  \\  
NH / ND  &  9  &   9 &  11 \\   
\NHHH / \fNHHH  &396 &  387   & 416   \\
\NHHD / \fNHHD  &  322&  336  &  376  \\
\NHHH / \NHHD  &11 & 18        &   33 \\
\NNHp / \NfNHp &425  &  423 &  417 \\
\NNHp / \fNNHp & 434 &   433  &  433 \\
\NNHp / \NNDp  & 8.7 & 8.6  & 9.7  \\
CN / \tCN   &  65 &  63  &  60 \\
CN / \CfiN  & 449&  445  &   438    \\
HCN / \tHCN  & 115 &  114 &   107 \\
HCN / \fHCN  & 467  & 453   &   458 \\
HCN / DCN  &  25 &   22 &  18 \\
HNC/ \tHNC &  120 &   121  & 118  \\
HNC/ \fHNC & 461 &   446 &   453   \\
HNC/ DNC &  19  & 16   &  12  \\
%DNC/ DN\tC &  5.1 & 12   &  13 &  &   &  30$^{+8}_{-5}$ (1) \\
 % {CO} /  {\tC}O & &     & 67 & 68 &      &       \\
%{CH} /  {\tC}H & &     & 78 & 101  &      &       \\
%{\HCOp} / {\HtCOp} & 17&     & 32 & 40 &     &     \\
{\HCOp} / {\HtCOp} &  58&  56 &51  \\
{\HCOp} / {\DCOp} &  8.6&  8.4 & 9.4  \\
\hline                    
  \hline                  
\end{tabular}
\end{table}

\begin{figure*}%[h]
   \centering
  \includegraphics[width=15cm]{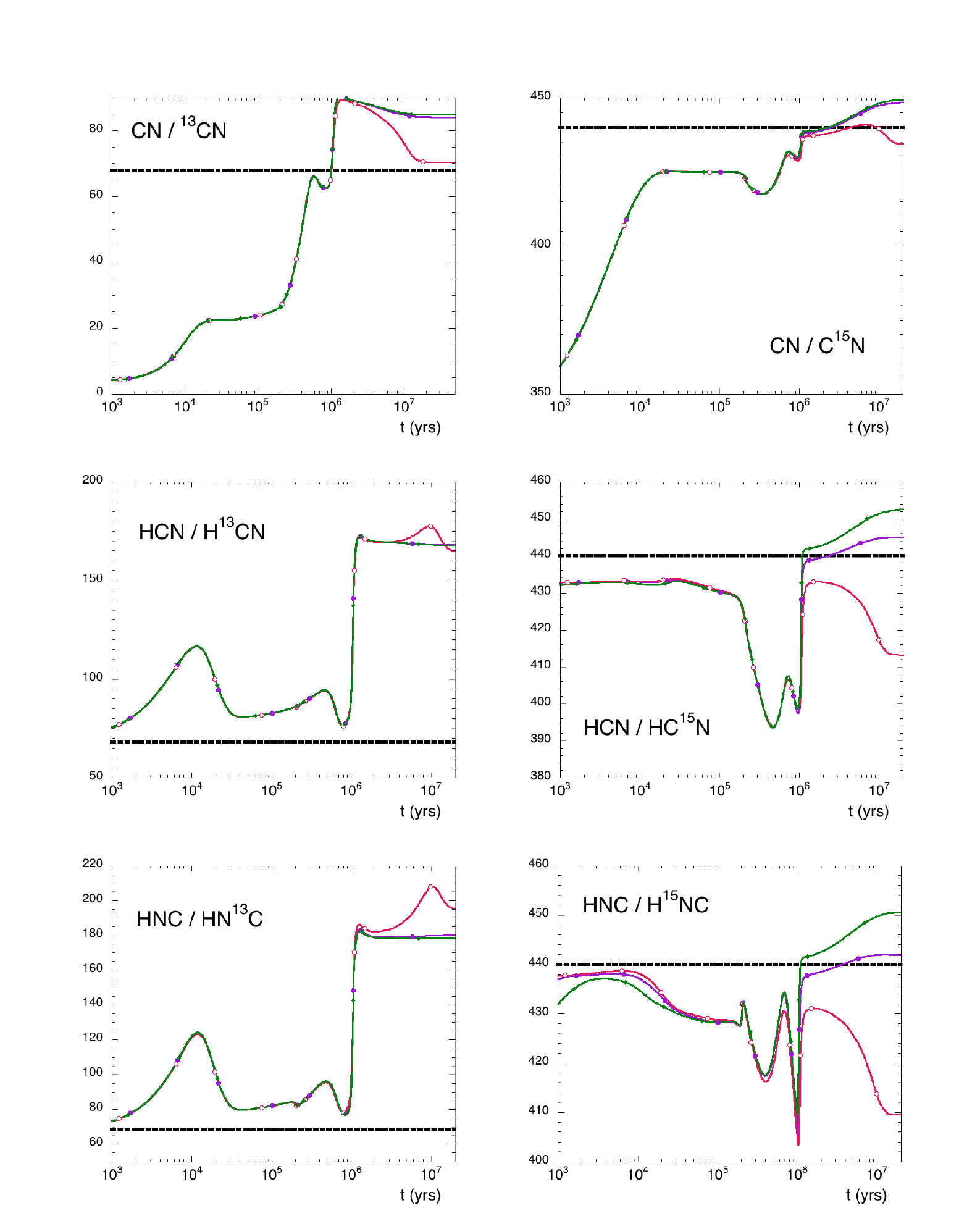}
   \caption{Time dependence of   fractionation ratios of CN, HCN, HNC in model (a) for 3 different OPR values. Black heavy dotted line: elemental ratio; green : OPR=10$^{-4}$, purple : OPR = 10$^{-3}$, red : OPR =10$^{-2}$}
      \label{fig:opa}%
    \end{figure*}

\begin{figure*}%[h]
   \centering
  \includegraphics[width=15cm]{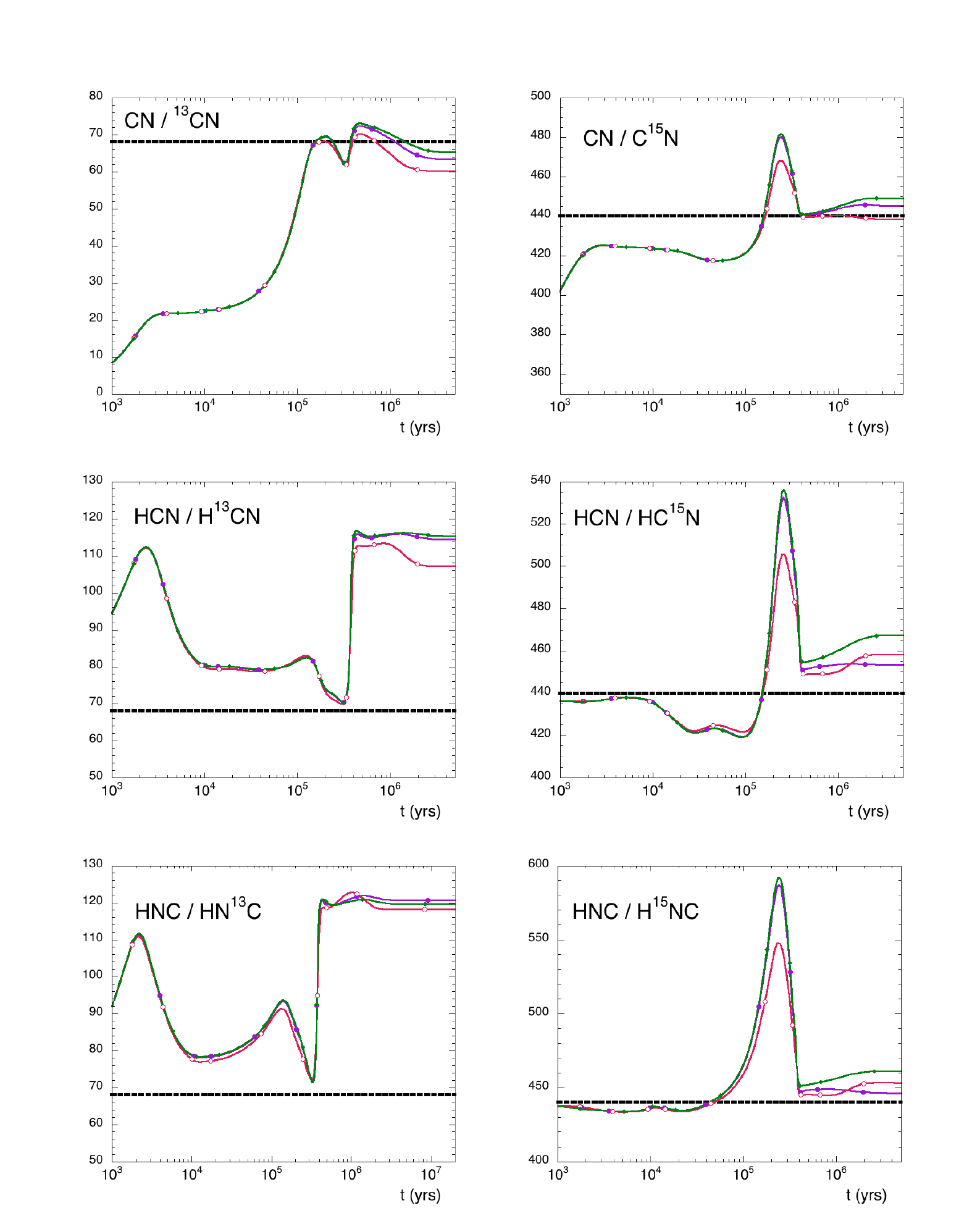}
   \caption{Time dependence of   fractionation ratios of CN, HCN, HNC in model (b) for 3 different OPR values. Black heavy dotted line: elemental ratio; green : OPR=10$^{-4}$, purple : OPR = 10$^{-3}$, red : OPR =10$^{-2}$}
      \label{fig:opb}%
    \end{figure*}

Tables \ref{tab:opa} and \ref{tab:opb} display the fractionation values for the three different assumed values of the OPR for models (a) and (b) %The variations are  {\bf{significative. }}%not spectacular although indicative. 
and Figures \ref{fig:opa} and \ref{fig:opb} display  the corresponding time evolutions. We see that the curves are 
almost superposable for times less than 1 Myr for model (a) and less than several 10$^5$ yrs for model (b). %and that the variations at steady state are moderate.%
 The variations are significative at steady state. We find that the competition between the destruction channels of  {\Np}
through its reactions with o-{\HH} and CO plays a major role.  If o-{\HH} is the most efficient destruction channel, which occurs typically for 
$\rm{OPR} >  200   \times  x_{CO} $ at 10K, where x$_{CO}$ is the fractional abundance of CO relative to \HH,
formation of  {\NHHHHp} proceeds efficiently. In the opposite case  {\Np} is mainly destroyed through  reaction with CO to yield  {\COp}, leading rapidly to {\HCOp}, and {\NOp} (which does not react with \HH). As the DR rate coefficients of polyatomic ions increase significantly  for larger polyatomic ions (the {\NHHHHp} + e$^-$ reaction is 3 times more rapid than {\HCOp} + e$^-$  and 20 times more rapid than {\HHHp} + e$^-$), these two different channels   impact the electron abundances and then affect many other species. 
 Moreover, the reaction between  {\NHHH} and {\HHHp}  leads to  {\NHHHHp}  + \HH. {\NHHHHp} then reacts mainly with electrons, recycling back to {\NHHH} which subsequently converts to {\NHHHHp}, amplifying the electron loss through DR. 
 These effects occur when CO becomes the main carbon reservoir, for sufficiently large evolution times. 
%An additional recycling mechanism even amplifies these effects as the reaction between {\NHHH} and {\HHHp} is more efficient than the DR of  {\HHHp} itself. 
As a result, the electron fraction
is found to decrease when  the OPR increases at large evolution times and at steady state.
%The impact of o-{\HH} on {\HHDp} is mitigated by the dissociative recombination of this ion and its reaction with other neutrals. In the chosen conditions, the decrease of the electronic fraction with increasing ROP is impacting the \HHDp / \HHHp and subsequenttly the other deuterium fractionation ratios such as HCN/DCN and HNC/DNC. 
The decreasing electron abundance with increasing OPR values acts to diminish the effect of all DR reactions
and as a result to increase the abundances of {\HHHp} and {\HHDp} since DR is the main destruction channel in our conditions. The abundance of {\HHDp} is even more enhanced through the {\HHHp} + HD reaction. Then  the abundances of other deuterated molecular ions produced through deuteron transfer  reactions of abundant neutrals  with {\HHDp} are increased as well, which impacts the subsequent formation of neutral deuterated compounds produced in the DR reactions. Eventually, the HCN/DCN and HNC/DNC ratios are shown to decrease for increasing OPR values.
This example demonstrates the complexity of the interplay between the different chemistries.

\section{Conclusions}
\label{sec:conclusion}
 We have built for the first time an isotopically substituted gas phase chemical network including D, {\tC} and {\fN} species and included it in a time dependent chemical model. Our model is based on a  careful  analysis of the possible gas phase mechanisms involved in carbon and nitrogen fractionation by scrutinizing the few available experimental studies and performing  DFT and ab-initio quantum calculations on  hypothetical reactions to check the possible presence of barriers in the reaction channels. One important result obtained is that the fractionation reaction of  {\fN} with {\NNHp} is unlikely, due to the presence of a barrier, in contrast to the previous hypothesis made by \cite{terzieva:00}. As a result, the modeled isotopic ratios involved in the isotopologues of \NNHp are  found to be very close to the elemental values and are similar to each other, in contradiction with observations towards L1544 \citep{bizzocchi:13}. The availability of new collisional rate coefficients
 for the {\NNHp - \HH} system \citep{lique:15} may however modify these conclusions. 
 We also discarded the {\fN} + {\HCNHp} and  {\fN}  + NO exchange reactions through similar arguments. Tentative reaction rate coefficients are also proposed for carbon fractionation reactions involving {\tCp} and {\tC} with CN and {\nCC}.
Additionally, we have explicitly considered the various isotopologues  involved in  N$^{+}$ +  {\HH} reaction, assuming that  the energy defect involved in the reactions of {\Np} with para-{\HH}  is a "real" endothermicity. This leads to a slight decline of the exponential term when {\fNp} reacts with {\HH} and with HD compared to {\Np}. This explains satisfactorily that {\fNHHD} is found to be more enriched in {\fN} than {\fNHHH} in the observations. 
Comparison between observations of nitriles and isonitriles and simulated values is much more questionable, as carbon and nitrogen chemistries are interdependent. Observed isotopic ratios are usually large and suffer from large error bars due to opacity effects in the main isotopologue and difficulties linked to nuclear spin effects. Whereas the various isotopologues follow a similar evolution, the isotopic ratios display significant variations due to slight shifts in the position of the maximum fractional abundances. 
A reasonable agreement {\bf{is}} obtained between the observed {\tC} / {\fN} ratios for most of the species in L134N and Barnard B1 and steady state model values. %may be misleading as 
Our model results  show a strong depletion in {\tC} and a near elemental ratio for  {\nN} / {\fN}, whereas  observations are usually  interpreted by assuming an elemental ratio for {\tC} containing species which leads to the incorrect assumption of {\fN} enrichment.  These considerations
are undeniably dependent on the chosen elemental abundances, and in particular to the assumed C/O ratio. 
 We additionally point out a somewhat unexpected effect of the ortho to para ratio of \HH, which affects significantly the fractional ionization, and consequently the level of deuterium fractionation  through the respective orders of magnitude of DR rate coefficients of polyatomic molecular ions. The importance of coupling C, O and N chemistries is emphasized. %
\section*{Acknowledgments} 
 We acknowledge support of PCMI (Programme National de Physique et Chimie du Milieu Interstellaire). This work has been  partially funded by the Agence Nationale de la
Recherche (ANR) research project IMOLABS (ANR-13-BS05-0008). 
%===================================================================================
% BIBLIOGRAPHY
%===================================================================================
%\bibliography{roueff}
\bibliography{roueff-n}
%\bibliography{roueff}
\bibliographystyle{aa}

\Online

\begin{appendix} 
\section{Exchange reactions}
We discuss below the reactions displayed in Table~\ref{tab:exch}.
\begin{itemize}
\item {\bf{$\NfiN + \NNHp$}}   $\rightleftharpoons$   {\bf{$ \NfNHp + \NN$}} \\
{\bf{$\NfiN + \NNHp$}}   $\rightleftharpoons$   {\bf{$ \fNNHp + \NN$}} \\
These reactions have been studied experimentally by \cite{adams:81} at 80K without differentiating between {\NfNHp} 
and  \fNNHp. The total rate is 4.6 $\times$ 10$^{-10}$  cm$^{3}$ s$^{-1}$. We thus take half this value for the forward reaction rate constant considering the symmetry factor and we assume no barrier for these rapid reactions.

We also introduce the  {\bf{$\NfiN + \fNNHp$}}   $\rightleftharpoons$ {\bf{$ \NfNHp + \NfiN$}} reaction for completeness.
\item {\bf{$\fNp + \NN$}}   $\rightleftharpoons$    {\bf{$ \Np + \NfiN$}}\\
This reaction involves adduct formation. The numerical constant is computed to reproduce the experimental value of  \cite{anicich:77} at room temperature.
 \item {\bf{$\fN  + \CNCp$}}   $\rightleftharpoons$     {\bf{$ \CfNCp  + \nN$}} \\
 No information is available for this reaction.The bimolecular exit channels {\Cp} + NCN and \nCCp / {\NN}  are both endothermic
 by 170 and 76 kJ/mol respectively. We performed DFT calculations (at the M06-2X/cc-pVTZ level) to explore the possibility of isotopic exchange. Direct N exchange is impossible as it would require simultaneous bond formation, rearrangement and bond breaking. However, isotopic exchange could take place through adduct formation. No barrier was found in the entrance channel for NCNC$^+$ formation (exothermic by 167 kJ/mol). The most favorable NCNC$^+$ isomerization pathway is cyclic. However the  position of the c-NC(N)C$^+$   transition state (TS) is highly uncertain and is found to be  below the entrance channel at M06-2X/cc-pVQZ level  (-14 kJ/mol) but slightly above the entrance level at the RCCSD(T)-F12/aug-cc-pVTZ level  (+19 kJ/mol). 
In any case, the TS position is close to the entrance level so that the NCNC$^+$ back dissociation is favored at room temperature. Isotopic exchange may however be  enhanced at low temperature and 
 may become the main exit channel. Considering a T$^{-1}$ temperature dependence of the adduct lifetime we tentatively suggest $k_f=3.8 \times 10^{-12} \times (T/300)^{-1}$    cm$^3$s$^{-1}$.
 \item  {\bf{$\fNp + \nNO$}}  $\rightleftharpoons$  {\bf{$ \Np + \fNO$}} \\
 \cite{terzieva:00} estimated the value of this rate coefficient  from the difference between the Langevin rate and the other measured exothermic reactions reported below. 
 \\
 \\
\begin{tabular}{llll} 
 % \fNp  + \nNO    &  $\rightarrow$ &    \fN  + \nNOp & $\Delta H_r$ =-509 kJ/mol & k=5.0 $\times$ 10$^{-10}$ cm$^3$ s$^{-1}$ \\
%                                &  $\rightarrow$ &   \fNNp  +  \nO &  $\Delta H_r$ =-213 kJ/mol     &   k=5.0 $\times$ 10$^{-11}$ cm$^{3}$ s$^{-1}$  \\
 %                               &  $\rightarrow$ &  \NfiN  +   \Op &  $\Delta H_r$ =-402 kJ/mol    &   no data              \\
%                                &  $\rightarrow$ &    \Np  +   \fNO & $\Delta H_r$ =-0.2 kJ/mol    &      no data              \\
  \fNp  + \nNO    &  $\rightarrow$ &    \fN  + {\nNOp} (k$_1$)& $\Delta H_r$ =-509 kJ/mol   \\
                                &  $\rightarrow$ &   \fNNp  +  {\nO} (k$_2$) &  $\Delta H_r$ =-213 kJ/mol     \\
                                &  $\rightarrow$ &  \NfiN  +   {\Op}   &  $\Delta H_r$ =-402 kJ/mol      \\
                                &  $\rightarrow$ &    \Np  +   {\fNO}  & $\Delta H_r$ =-0.2 kJ/mol       \\
\end{tabular} 
 \\
 \\
 k$_1$ = 5.0 $\times$ 10$^{-10}$ cm$^3$~s$^{-1}$ and   k$_2$ = 5.0 $\times$ 10$^{-11}$ cm$^{3}$~s$^{-1}$. No data are
 available for the other two reactions.
 We find a barrier for NON$^+$ adduct formation (261kJ/mol at the M06-2X/cc-pVTZ level) and neglect this exchange reaction. 
 \item  {\bf{$\fN  + \NNHp$}}   $\rightleftharpoons$      {\bf{$\nN  +   \NfNHp / \fNNHp$}} \\
 No information is available for these reactions, which were reported as crucial by \cite{rodgers:08a}. 
 We performed DFT calculations (at the M06-2X/cc-pVTZ level) which showed that isotopic exchange through the addition
 elimination mechanism  is 
 impossible as  the NN(H)N$^+$ and NNNH$^+$ ions are metastable, respectively 73 and 332 kJ/mole above the 
 \nN  +  \NNHp energy.
    \begin{figure*}%[h]
   \centering
  \includegraphics[width=14cm]{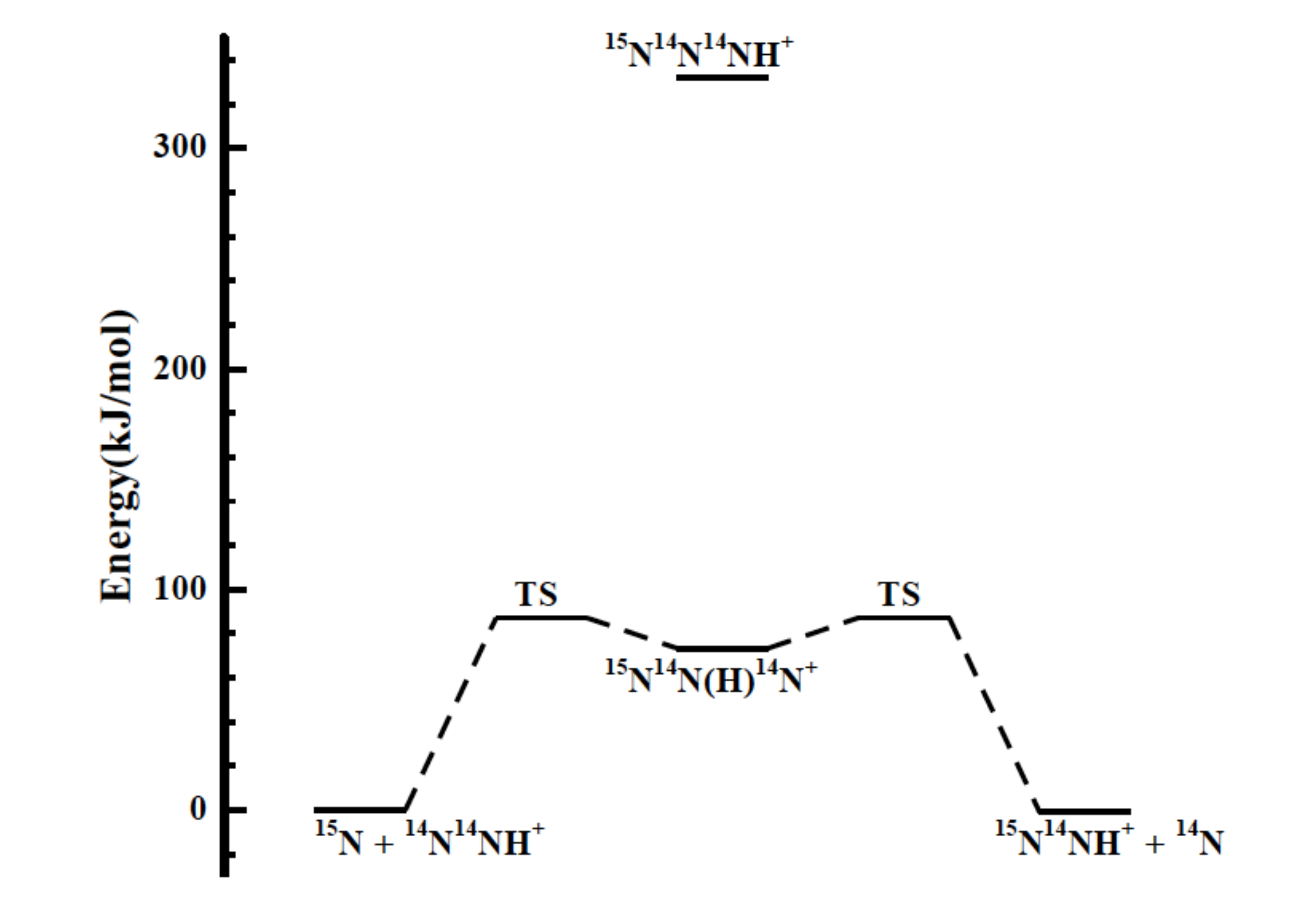}
   \caption{Energy diagram of the NNNH$^+$ system.}
              \label{fig:nnnhp}%
    \end{figure*}
Figure \ref{fig:nnnhp} displays the various possibilities. In addition, direct nitrogen exchange through NN(H)...N$^+$ $\rightarrow$ N...N(H)N$^+$ presents a barrier equal to +87 kJ/mol.
We did not find any N$_3$H$^+$ configuration with an energy lower than that of the reactants. We then neglect this reaction.
\item {\bf{$ \fNNHp + \nH  $}}   $\rightleftharpoons$     {\bf{$\nH  +   \NfNHp $}} \\
Following a similar suggestion for explaining abundance anomalies of the  $^{13}$C species of CCH \citep{sakai:10}, we consider the above reaction for which no information is available. 
We performed various DFT calculations  at the M06-2X/cc-pVTZ level. The HNNH$^+$ ion is 109 kJ/mole more stable than H + {\NNHp}
but the addition reaction shows a barrier of +12kJ/mole. The intramolecular isomerization \fNNHp $\rightarrow$ {\NfNHp} displays also a barrier of +172 kJ/mole. Then direct isomerization  through tunneling is expected to be very slow and we neglect that reaction \footnote{we note in addition that {\NNHp} formation from {\NN} + {\HHHp} is exothermic by only 72 kJ/mole, so that any {\fNNHp} produced will not isomerize into {\NfNHp}  and vice versa.}.

 \item{\bf{$\fN  + \HCNHp$}}  $\rightleftharpoons$    {\bf{$\nN  +   \HCfNHp $}} \\
 This reaction was also suggested by \cite{terzieva:00} despite the lack of any relevant information. We performed DFT calculations to explore the possibility of isotopic exchange as shown in Figure \ref{fig:hcnnhp}.
    \begin{figure*}%[h]
   \centering
  \includegraphics[width=14cm]{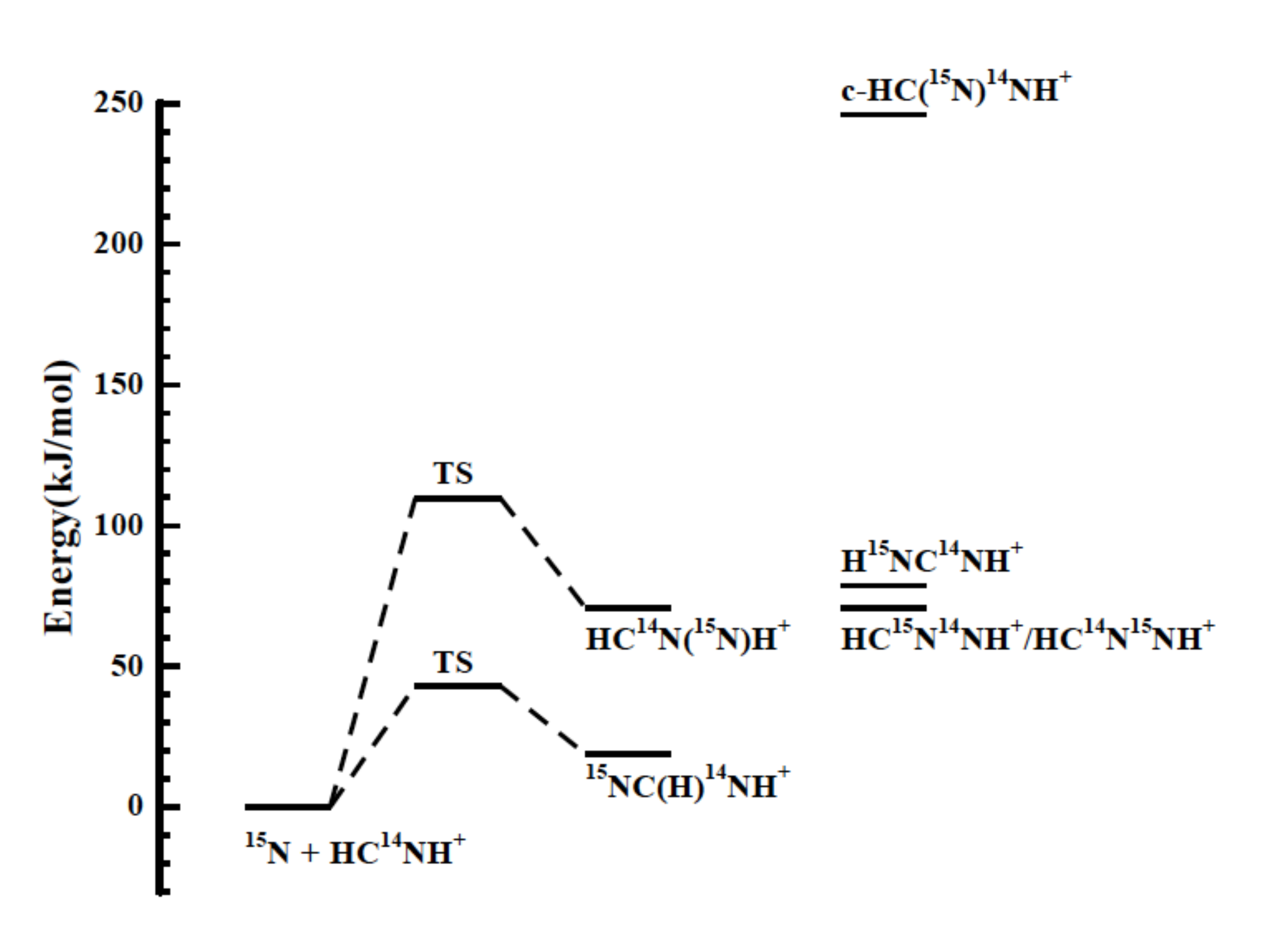}
   \caption{Energy diagram of the HCNNH$^+$ system.}
              \label{fig:hcnnhp}%
    \end{figure*}
Direct N exchange is found to be impossible as it requires simultaneous bond formation, rearrangement and bond breaking 
so that isotopic exchange involves adduct formation.  Attack by atomic nitrogen on either side of the {\HCNHp}  molecular ion leads to a metastable system through high energy transition states.  We did not consider transition states leading to N addition on the C=N bond of {\HCNHp}  nor did we examine N insertion into the N-H or C-H bonds, as all these pathways lead to species located above the reagent energy level.
We did not find any CN$_2$H$_2$$^+$ species lying below the reagent energy level and we consider that this reaction cannot take place.
\item {\bf{$\fN  + \nCN$}}   $\rightleftharpoons$     {\bf{$ \nN  + \CfiN$}} \\
The N($^4S_u$) + CN ($^2\Sigma^+$) reaction leads to $^{3,5}\Sigma^-$  surfaces in C$_{\infty V}$ symmetry and 
$^{3,5}$A$^{\prime\prime}$ surfaces in Cs
symmetry. The quintuplet surface is repulsive at the MRCI+Q/aug-ccpVTZ level and the
$^5$NCN intermediate is above the N + CN level. Considering only the triplet surface, the only
barrierless reaction is attack on the carbon atom leading to the ground state $^3$NCN intermediate \citep{daranlot:12,ma:12}
. The main exit channel is C + {\NN} after isomerization of the NCN
intermediate through a tight TS and then back dissociation may be important and isotope
exchange possible.
In the nominal model we neglect this reaction but some tests have been performed to estimate its potential role.
The upper limit of the isotope exchange rate constant is equal to the capture rate constant minus the N + CN $\rightarrow$ C + {\NN} 
rate constant.
That value is notably smaller than the capture one at low temperature. We thus propose the upper limit value of
 2.0 $\times$ 10$^{-10}$ $\times$ (T/300)$^{1/6}$ $\frac{1}{1+exp(-22.9/T)}$ cm$^3$ s$^{-1}$ for the forward rate constant. 
\item {\bf{$\fN  + \CCN$}}   $\rightleftharpoons$   {\bf{$ \nN  + \CCfN $}} \\
    \begin{figure*}%[h]
   \centering
  \includegraphics[width=14cm]{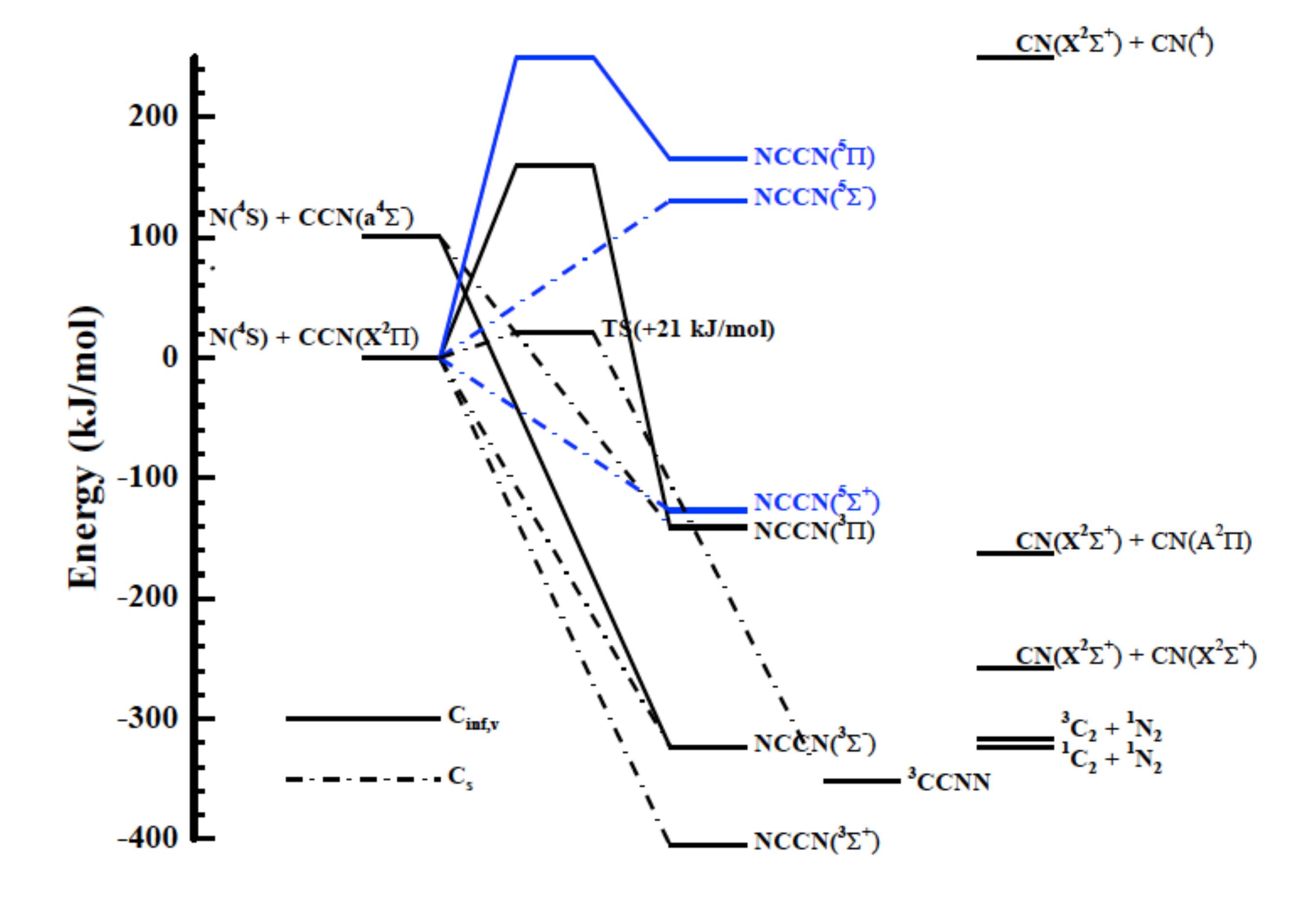}
   \caption{Energy diagram of the CCNN  system.}
              \label{fig:cncn}%
    \end{figure*}
The N($^4$S) + {\CCN} ($^2\Pi$) reaction leads to $^{3,5}\Pi$  surfaces in  C$_{\infty V}$ symmetry and 
$^{3,5}$A$^{\prime}$ + $^{3,5}$A$^{\prime\prime}$ 
surfaces in Cs symmetry. 
Figure \ref{fig:cncn} displays the positions of the state energies   linked to {N + \CCN} which are calculated at the 
MRCI+Q/aug-cc-pVTZ level (10 e- in 10 OM with the geometry fully
optimized at the CASCCF level or (14 e- in 14 OM for the CASCCF and 14e- in 12 OM for
the MRCI calculation with non relaxed geometry).
The triplet surface pathway leads to a  NCCN adduct  in a triplet state, corresponding to an excited
state of the very stable NCCN  linear molecule, located -393 kJ/mol below the N($^4$S) + {\CCN} ($^2\Pi$) 
level.  No barrier is present in the entrance valley so that this NCCN adduct very likely leads
 to CN (in a doublet state) + CN (in a doublet state) products. The occurrence of a  small
exit barrier cannot be excluded but its energy should be much lower   than that of the $ \nN  + \CCfN $ entrance channel. 
 Triplet surfaces thus cannot lead to isotopic exchanges. 
%Considering the exothermicity of the N + {\CCN}   $\rightarrow$ CN + CN reaction (-293
%kJ/mol) and the absence of a tight TS for passage to bimolecular products, it is unlikely that
%triplet surfaces lead to substantial isotopic exchange as $^3$NCCN back dissociation should be
%negligible.
However  the quintuplet surface deserves specific attention as the  % this reaction is very interesting because there is the possibility to form 
NCCN($^5 \Sigma^+$)
 adduct is found at an energy  of -183 kJ/mol below that of  N($^4$S) +  {\CCN} ($^2\Pi$) level  (at the MRCI+Q, RCCSD(T) and DFT level of calculations).
 However, no  exothermic
bimolecular exit channel is available on this quintuplet surface. 
 We thus conclude that the quintuplet surface could lead to isotopic exchange and consider finally the two following possibilities 
for this reaction involving the very reactive {\CCN} radical \citep{wang:05,wang:06}, i.e.
the triplet channel :
\begin{eqnarray} 
 \fN(^{4}S) + \CCN (^2\Pi)  \rightarrow  \fNCCN (^3\Sigma^+,^3\Sigma^-)    \\
                                            \rightarrow     \CfiN (^2\Sigma^+,^2\Pi)  + \nCN (^2\Sigma^+,^2\Pi)  
 \end{eqnarray}
and the quintuplet channel :
 $$ \fN(^{4}S) + \CCN (^2\Pi) \rightarrow  \fNCCN (^5\Sigma^+)  \rightarrow  \fN(^{4}S) + \CCN (^2\Pi) 
  $$ 
 $$ \fN(^{4}S) + \CCN (^2\Pi) \rightarrow  \fNCCN (^5\Sigma^+)  \rightarrow  \nN(^{4}S) + \CCfN (^2\Pi) 
  $$ 
 
%Moreover,   {\CCN} ($^2\Pi$) presents a very high reactivity \citep{wang:05,wang:06},
 %similar to the CH radical due to the presence of free doublet electrons and
%a low energy vacant molecular orbital both localized on the carbon atom furthest from N in
%the {\CCN} molecule.
%Figure \ref{fig:cncn} displays the positions of the state energies   linked to {N + \CCN} which are calculated at the 
%MRCI+Q/aug-cc-pVTZ level (10 e- in 10 OM with the geometry fully
%optimized at the CASCCF level or (14 e- in 14 OM for the CASCCF and 14e- in 12 OM for
%the MRCI calculation with non relaxed geometry).
%The potential approach in C$_S$ symmetry  for the  $^5$A$^\prime$ surface displays a very small submerged barrier. 
 %
%
%Considering the reduced active space used in this work and the fact that the MRCI + Q method overestimates the repulsion energy, there is very likely to be no barrier and we conclude that isotopic exchange is possible.
%
\item {\bf{$\fN   + \nNO$}}   $\rightleftharpoons$  {\bf{$\nN  + \fNO$}} \\
There is a barrier for NON adduct formation. The isotope exchange rate is calculated to be
very low \citep{gamallo:10}  as the main exit channel is {\NN} + O. Moreover the quintuplet
surfaces are repulsive. We neglect this reaction.
\item  {\bf{$\tCp  + \nCO$}}  $\rightleftharpoons$  {\bf{$ \Cp  + \tCO$}}\\
This reaction was first mentioned by \cite{watson:76} and has been experimentally studied in detail by \cite{smith:80}
in the 80-500~K temperature range. Their data can be fitted \citep{liszt:07}. We introduce a new formula allowing us to describe the full temperature range with a single formula.
\item {\bf{$\tCO   + \HCOp$}}  $\rightleftharpoons$   {\bf{$ \nCO   + \HtCOp$}} \\
This reaction has also been studied experimentally by \cite{smith:80}. The exothermicity of the reaction has been reported to be 9 K by \cite{langer:84} from theoretical studies whereas \cite{smith:80} proposed a value of 12 $\pm$ 5 K. However, a later study by \cite{lohr:98}
leads to a value of 17.4 K. \cite{mladenovic:14} have reconsidered this reaction and have confirmed the value of \cite{lohr:98}. We include this exothermicity value in the present work. 
Experimental points are fitted through a power law as given in Table \ref{tab:exch}.
\item  {\bf{$\tCp  + \nCN$}}   $\rightleftharpoons$  {\bf{$ \Cp  + \tCN$}} \\
There are no bimolecular exit channels for this reaction and two different pathways lead to isotopic exchange. In a direct reaction,  \tCp + NC $\rightarrow$ $^{13}$CNC$^+$ $\rightarrow$ {\tCN}  + \Cp whereas the indirect pathway involves \tCp + CN $\rightarrow$ 
$^{13}$CCN$^+$ $\rightarrow$ \tCN + \Cp. There is no barrier in the entrance valley for both cases at the M06-2X/cc-pVTZ level. 
Moreover, the cyclic transition state from CCN$^+$ and CNC$^+$ is located at -436kJ/mol below the reactant energy, leading to fast isomerization. The corresponding capture rate constant is 3.82 $\times 10 ^{-9}  \times (T/300)^{-0.40}$.
\item  {\bf{$\tC   + \nCN$}}  $\rightleftharpoons$    {\bf{$ \dC   + \tCN$}}
In a similar way to the ionic case, no exothermic bimolecular exit channels are available for this reaction and two reaction pathways occur for possible isotopic exchange.
\tC  + NC $\rightarrow$ $^{13}$CNC  $\rightarrow$ \tCN + {\dC}  holds for the direct process and 
 \tC  + CN   $\rightarrow$    $^{13}$CCN $\rightarrow$ c-$^{13}$CNC $\rightarrow$ {\tCN}  + {\dC} describe the indirect exchange mechanism. No barrier is found in the entrance channel and the transition state from CCN to c-CCN is low (-298 kJ/mol below the reactant energy), in good agreement with \cite{mebel:02} so that isomerization is expected to be fast. The capture rate is computed as 3.0 $\times 10^{-10}$ cm$^3$ s$^{-1}$.
\item  {\bf{$\tC   + \nHCN$}}  $\rightleftharpoons$    {\bf{$ \dC   + \tHCN$}} \\
There are no exothermic bimolecular exit channels for this reaction and there is no barrier for
HCNC formation but isotopic exchange requires isomerization through a TS located close to
the reactant level, involving a TS located  at -16 kJ/mol at the M06-2X/cc-pVQZ level but +34
kJ/mol at the RCCSD(T)-F12-aug-cc-pVQZ level. Calculation of the rate constant for exchange
is complex and similar to the {\fN} + {\CNCp} case.
In the nominal model we neglect this reaction.
We estimate nevertheless an upper limit for the rate constant by using a 1/T temperature dependence and assuming that the isotopic exchange rate constant balances the back dissociation at very low temperature (3K). 
\item {\bf{$\tC   + \nCC$}}   $\rightleftharpoons$   {\bf{$ \dC   + \tCC$}}  \\
There are no exothermic product channels for this reaction and there is very likely to be no barrier. The capture rate constant is equal to 3.0 $\times 10^{-10}$ cm$^3$ s$^{-1}$.  
\item {\bf{$ \tCH + \nCO $}}   $\rightleftharpoons$   {\bf{$ \tCO   + \nCH$}}   \\ 
This reaction could be  an additional possibility to enhance {\tCO} as no barrier has been found in the entrance valley \citep{lepicard:98}. The high pressure CH + CO $\rightarrow$ HCCO  association reaction rate constant is equal to 3 $\times 10^{-11} \times (T/300)^{-0.9}$
between 53 and 294~K. The exchange rate has been measured to be $\sim \times  10 ^{-12}$  \citep{taatjes:97} at room temperature. 
 This value represents {1\%} of the association reaction rate constant at high pressure, which is explained by a transition state localized at  6kJ/mole above the reactants energy as computed at the M06-2X/cc-pVTZ level, in good agreement with \cite{sattelmeyer:04}. We do not include this reaction in our models given the high TS which should make this exchange process negligible at low temperature.
\end{itemize}

\section{ZPE values}
We revisit the ZPE values in light of several recent studies. 
 \onltab{
 \begin{table}[ht]
    \caption {Atomic Masses in amu. from NIST }
         \label{tab:atom}
    \begin{center}
      \begin{tabular}{lc  } 
        \hline \hline
  Atom  &  Mass      \\
  \hline
 H  &1.007825032 07   \\
 D  &              2.014 101 777 8 \\
 N  &14.003074004   \\
 $^{15}$N   &15.000108 898  \\
C  &12.00   \\
 $^{13}$C   &13.003354 838  \\
O &15.994 914 619 6   \\
 $^{18}$O   &17.999 161 0   \\
 e   & 0.00054858     \\
  \hline \hline
\end{tabular}
 \end{center}
 \end{table}
 }
 We recall in Table \ref{tab:atom} the atomic masses of various isotopes which may be used to derive spectroscopic constants 
 of isotopic molecules, as found in basic molecular spectroscopy textbooks \citep{herzberg:45,herzberg:89}.   We only refer to the first order expansion terms for the purpose of computing ZPEs.

\subsubsection{Diatomic Molecules}
For diatomic molecules, energy levels are expressed  as a Dunham expansion or equivalently as a sum of harmonic + anharmonicity correction factors. The following expression is obtained for the ZPE, corresponding to $v=0$. 
\begin{equation}
ZPE= \frac{1}{2} Y_{10}  +  \frac{1}{4} Y_{20} = \frac{1}{2} \omega_0 -  \frac{1}{4}  \omega_ex_e
\end{equation}
$\omega_0$ is the harmonic contribution of the vibrational frequency and $\omega_ex_e$ represents the anharmonic contribution. The reduced mass $\mu$ dependence  of $\omega_0$ and $\omega_ex_e$ are 1/$\sqrt{\mu}$ and 1/$\mu$ respectively. The label  $computed$ in Table \ref{tab:diatom} indicates the use of this property to compute the spectroscopic constants, in the absence of other information.
\onltab{
 \begin{table}[h]
    \caption {Spectroscopic constants  of diatomic  molecules in {\cm} and differences of ZPE with respect to the main isotopologue.}
         \label{tab:diatom}
    \begin{center}
      \begin{tabular}{lcccc c} 
        \hline \hline
  Molecule  &  Ref & $\omega_e$    &   $\omega_e$x$_e$  & ZPE & $\Delta$(ZPE) (K) \\
  \hline
  \NN  &[1] & 2358.53   &    14.30       &1175.7&  - \\
 \NfiN  &[1]   &2319.01   &    13.83      &1156.05&  28.3 \\
\fiNN  &[1]  &2278.80   &    13.35    &1136.06&   57.0 \\
 \nNH  &  [2]  &   3282.2       &     78.3         & 1621.5 & -   \\
 \fNH  & $computed$  &     3274.9     &  77.95    &    1617.9& 5.2   \\
 \nND  & [3]& 2399.0         &    42.0   &    1189.5& 621.6 \\
 \fND & $computed$  &     2389.0    &  41.6   &    1183.6 & 630  \\
\NHp &[4] &   3047.58    & 72.19       &  1505.7   & \\
\fNHp &[4] &   3040.77    &    71.87    &   1502.4 &  4.8  \\
\NDp &[4] &   2226.93    &   38.55    &  1103.8  &  578  \\
\fNDp &$computed$ &   2217.60    &    38.23      & 1099.2 &  585  \\
\nCN  &[5]&2068.68   &    13.12  &    1031.1 &  - \\
\tCN  &[6]&2025.25   &    12.57  &       1009.5& 31.1 \\
\CfiN &[7] &   2036.70 &  12.70   &      1015.2 & 22.9 \\
\nNO &[8]&   1904.11 &   14.08  &   948.5 &-  \\  
\fNO &[9]&   1870.1 &   13.585   &    931.7 &  24.3 \\ 
 \nCC  &[10] & 1853.5 &   12.755  &    923.98&  - \\
  \tCC  &[10] & 1817.3  &  12.1  &    905.61 & 26.4 \\
 \nCO  &[11] &  2169.8&  13.3    &  1081.6    &  - \\
  \tCO  &[11] & 2121.4 & 12.7      & 1057.5 &    34.7  \\
 \nCH  &[12] &  2860.75&  64.44   &  1414.3    &  - \\
  \tCH  &[13] & 2852.15 & 64.04      & 1410.1 &    6.0  \\
  \nCD  &[14] &  2101.05 & 34.7    &  1042.1   & 535.5 \\
\hline \hline
\end{tabular}
 \end{center}
 [1] \citealt{leroy:06}, [2] \citealt{ram:10c}, [3]  \citealt{dore:11}, [4] \citealt{colin:89}, [5]   \citealt{ram:10a}, [6] \citealt{ram:12},
 [7] \citealt{colin:12}, [8]  \citealt{henry:78},    [9] \citealt{danielak:97},  [10]  \citealt{zhang:11}, [11]  \citealt{guelachvili:83},
 [12] \citealt{zachwieja:95},  [13] \citealt{zachwieja:97},  [14] \citealt{zachwieja:12}
 \end{table}
 }

\subsubsection{Polyatomic molecules}

For polyatomic molecules, the following expression extends the diatomic formulae where the different vibrational degrees of freedom are included:
\begin{equation}
G_0=\sum _i \omega_i\frac{d_i}{2} + \sum_i \sum_j x_{ij}\frac{d_i}{2} \frac{d_j}{2}
\end{equation}
$\omega_i$ refers to the harmonic frequencies, $d_i$ is the corresponding degeneracy and  x$_{ij}$ stands for the anharmonic terms. The sum is performed over the number of vibrational modes. In the case of polyatomic linear molecules, the number of vibrational modes is 3N-5 whereas it is 3N-6 in the general case where N is the number of nuclei in the molecule.
\onltab{
\begin{table}[h]
    \caption {Spectroscopic constants  of triatomic  molecules in {\cm} and differences of ZPE with respect to the main isotopologue.}
             \label{tab:triatom}
     \begin{center}
      \begin{tabular}{lcccccc} 
        \hline \hline
   Molecule  &Ref   &  $\omega_1$    & $\omega_2$  &   $\omega_3$    &  ZPE &$\Delta$(ZPE) (K) \\
symmetry   &           &  $\Sigma^+$           & $\Pi$                  & $\Sigma^+$        &                \\
\hline   
  \nHCN  & [1]& 3443.1  & 727.0     & 2127.4  &       3512.25          &      -    \\
\tHCN    & [1]& 3424.0   & 720.6   & 2093.0     &        3478.6          &   48.4     \\ 
\fHCN    & [1]& 3441.7  & 725.9    & 2094.0       &       3493.75           &     26.6   \\ 
\tfHCN    &[1]& 3422.6  & 719.5    & 2058.6       &       3460.1          &  75.0  \\ 
 DCN  & [2] &  2702.5  & 579.7     & 1952.3  &       2883.9      &     904.1  \\
  CCN  & [3] &  1967.2  & 322.2     & 1058.3&     1794.20     &  -  \\
 $^{13}$CCN  &   [3]   &            &         &     &     1778.19   & 23.0  \\
 C$^{13}$CN  &   [3]  &        &          &     &     1761.11    & 47.6 \\
 CC$^{15}$N  &  [3]  &       &         &        &     1775.65    &  26.7  \\
 \nHNC & [4]& 3819.9  & 463.8     & 2064.3  &        3369.9            &   -    \\
  \tHNC & [4] &   -    & 463.5     &    - &                      &     \\
  \fHNC & [4]  &   -  & 461.5     &   -    &                       &    \\
  \NNHp & [5] & 3398.6    &695.5    & 2297.7      &3508.6   &  -    \\
    \fNNHp &[5] & 3396.5    &694.4     & 2259.3      &    3487.5                &  30.4     \\
  \NfNHp & [5] & 3382.4    &690.7     & 2268.7      &    3481.8             &    38.5      \\
 \ffNNHp & [5]   & 3380.7    &699.6     & 2229.3      &         3460.4      &     69.3       \\
 \NNDp & [5] & 2719.8    &550.4     & 2065.9      &      2916.2               &  852    \\ 
 \NfNDp  & [5] & 2680.4   &544.4   & 2058.8        &      2888.4  &         40.0       \\
 \fNNDp & [5] & 2703.8    &549.0     & 2041.0       &        2895.2           &      30.2        \\
 \CNCp &   [6] &  1335   &   165   &   2040 &     1852.5 &    -   \\
 \CfNCp   &  [6]  &  1334  &  161   &    1997  &     1826   &    38.1   \\
 \HCOp & [7], [9]& 3229.6   &  832.5 &  2195.8 &3524.60 &  - \\  
 \HtCOp & [8],[9] &  3206.1 &  824.9 &  2160.6 &   3488.24 & 52.3 \\
 \DCOp & [7],[9] & 2645.9 & 666.5  & 1928.0  &     2944.22 &  835.1 \\
  \hline \hline
\end{tabular}
 \end{center}
     \label{tab:tri}
     [1] \citet{maki:00},  [2] \citealt{mollmann:02}, [3] \citealt{grant:11,mitrushcenkov:14}, [4] \citealt{maki:01}, [5] \citealt{huang:10}, [6] \citealt{jensen:88}, [7] \citealt{martin:93}, [8] private communication by T. Lee, based on   second-order vibrational perturbation theory using
a quartic force field determined at the CCSD(T)/cc-pVTZ level of electronic
structure theory, as in [7], [9] ZPEs are from \cite{mladenovic:14}.
 \end{table}
 }

\onltab{
\begin{table}[h]
    \caption {Spectroscopic constants  of tetratomic nitrogen molecules in {\cm} and differences of ZPE with respect to the main isotopologue. }
    \begin{center}
      \begin{tabular}{lcccccccc} 
        \hline \hline
   Molecule  &Ref   &  $\omega_1$    & $\omega_2$  &   $\omega_3$    &  $\omega_4$ & $\omega_5$ &ZPE &$ \Delta$(ZPE) (K) \\
 symmetry   &           &  $\Sigma$         & $\Sigma$      &   $\Sigma$         &       $\Pi$           & $\Pi$      &         \\
\hline   
\HCNHp  & [1]& 3484.2 & 3187.3 & 2159.9 & 804.8& 649.4  &   5987.3 &  -\\
\HtCNHp  & [1]& 3479.5 & 3170.1& 2126.1 & 799.8& 647.4  &   5952.0& 50.8 \\
\HCfNHp  & [2]& 3469.3& 3187.1 & 2133.3 & 804.7& 645.1  &  5961.5 & 37.1\\
\HCNDp  & [1]& 3221.9 & 2681.1 & 2025.9& 794.3& 513.1  &  5364.2 & 896.5\\
\DCNHp  & [1]& 3474.0 & 2601.0 & 1934.8 & 664.7 & 622.8  &  5388.0 & 862.3\\
    \hline \hline
   Molecule  &Ref   &  $\omega_1$    & $\omega_2$  &   $\omega_3$    &  $\omega_4$ &   & ZPE & $\Delta$(ZPE) (K) \\
    symmetry       &           &   A$_1$    &  A$_1$    &   E    &     E     &         &     &    \\
                          \hline
         \NHHH   &  [3]    &      3336.1    &    932.4  &   3443.6  &  1626.3   &    & 7429.2   &-          \\           
         \fNHHH   &    [3]    &   3333.3       &  928.5    & 3435.1    &  1623.2   &       & 7413.0  &  23.3     \\           
\hline   \hline
\end{tabular}
 \end{center}
     \label{tab:tri}
     [1] \citealt{brites:12}, [2] Brites, private communication, [3] \citealt{huang:11}
 \end{table}
 }
 \end{appendix}
 \end{document}